\documentclass[fleqn,usenatbib]{mnras}

\usepackage{newtxtext,newtxmath}

\usepackage[T1]{fontenc}

\DeclareRobustCommand{\VAN}[3]{#2}
\let\VANthebibliography\thebibliography
\def\thebibliography{\DeclareRobustCommand{\VAN}[3]{##3}\VANthebibliography}

\usepackage{graphicx}	
\usepackage{amsmath}	
\usepackage[flushleft]{threeparttable} 
\usepackage{comment}
\usepackage{caption}
\usepackage{subcaption}
\usepackage{rotating}
\usepackage{siunitx}
\usepackage{lscape}
\usepackage{mathtools}
\usepackage{siunitx}

\newcommand{\msun}{\mbox{$M_{\odot}$}}
\newcommand{\mch}{\mbox{M$_{\rm ch}$}}

\newcommand{\kms}{\mbox{$\rm{km}\,s^{-1}$}}

\newcommand{\nick}{\mbox{$^{56}$Ni}}
\newcommand{\cob}{\mbox{$^{56}$Co}}
\newcommand{\fe}{\mbox{$^{56}$Fe}}

\newcommand{\FeIF}{[Fe~{\sc i}]}
\newcommand{\FeIIF}{[Fe~{\sc ii}]}
\newcommand{\FeIIIF}{[Fe~{\sc iii}]}

\newcommand{\CoIIC}{[Co~{\sc ii}]}
\newcommand{\CoIIIC}{[Co~{\sc iii}]}

\setcitestyle{citesep={,}}

\title[A photometric study of the NIR plateau in SNe Ia]{Photometric study of the late-time near-infrared plateau in Type Ia supernovae}

\author[M. Deckers et al.]{
M. Deckers,$^{1}$\thanks{E-mail: deckersm@tcd.ie} O. Graur,$^{2,3}$ K. Maguire,$^{1}$ L. Shingles,$^{4}$ S. J. Brennan,$^{5, 6}$ J.~P. Anderson,$^{7, 8}$
\newauthor
J. Burke,$^{9, 10}$ T.-W. Chen, $^{11, 12}$ L. Galbany,$^{13,14}$ M. J. P. Grayling,$^{15,16}$ C. P. Guti\'errez,$^{17, 18}$ 
\newauthor
L. Harvey,$^{1}$ D. Hiramatsu,$^{19, 20}$ D. A. Howell,$^{9, 10}$ C. Inserra,$^{21}$ T. Killestein,$^{22}$
C. McCully,$^{10}$
\newauthor
T. E. Müller-Bravo,$^{13,14}$ M. Nicholl,$^{23}$ M. Newsome,$^{9, 10}$ E. Padilla Gonzalez,$^{9, 10}$ C. Pellegrino,$^{9, 10}$
\newauthor
 G. Terreran,$^{9, 10}$ J. H. Terwel,$^{1,24}$ M. Toy,$^{15}$ D. R. Young$^{25}$\\
$^1$School of Physics, Trinity College Dublin, College Green, Dublin 2, Ireland\\
$^2$Institute of Cosmology and Gravitation, University of Portsmouth, Portsmouth, PO1 3FX, UK\\
$^3$Department of Astrophysics, American Museum of Natural History, Central Park West and 79th Street, New York, NY 10024, USA\\
$^4$GSI Helmholtzzentrum für Schwerionenforschung, Planckstraße 1, 64291 Darmstadt, Germany\\
$^5$School of Physics, University College Dublin, Belfield, Dublin 4, Ireland\\
$^{6}$Department of Astronomy, The Oskar Klein Center, Stockholm University, AlbaNova, 106 91 Stockholm, Sweden \\
$^{7}$European Southern Observatory, Alonso de Córdova 3107, Casilla 19, Santiago, Chile\\
$^{8}$Millennium Institute of Astrophysics MAS, Nuncio Monsenor Sotero Sanz 100, Off. 104, Providencia, Santiago, Chile \\
$^{9}$Las Cumbres Observatory, 6740 Cortona Dr. Suite 102, Goleta, CA, 93117, USA\\
$^{10}$Department of Physics, University of California, Santa Barbara, CA 93106-9530, USA\\
$^{11}$Technische Universit{\"a}t M{\"u}nchen, TUM School of Natural Sciences, Physik-Department, James-Franck-Stra{\ss}e 1, 85748 Garching, Germany\\
$^{12}$Max-Planck-Institut f{\"u}r Astrophysik, Karl-Schwarzschild Stra{\ss}e 1, 85748 Garching, Germany \\
$^{13}$Institute of Space Sciences (ICE, CSIC), Campus UAB, Carrer de Can Magrans, s/n, E-08193 Barcelona, Spain\\
$^{14}$Institut d’Estudis Espacials de Catalunya (IEEC), E-08034 Barcelona, Spain\\
$^{15}$School of Physics and Astronomy, University of Southampton, Southampton, SO17 1BJ, UK\\
$^{16}$Institute of Astronomy and Kavli Institute for Cosmology, Madingley Road, Cambridge, CB3 0HA, UK\\
$^{17}$Finnish Centre for Astronomy with ESO (FINCA), FI-20014 University of Turku, Finland\\
$^{18}$Tuorla Observatory, Department of Physics and Astronomy, FI-20014 University of Turku, Finland\\
$^{19}$Center for Astrophysics, Harvard \& Smithsonian, 60 Garden Street, Cambridge, MA 02138-1516, USA\\
$^{20}$The NSF AI Institute for Artificial Intelligence and Fundamental Interactions, USA\\
$^{21}$Cardiff Hub for Astrophysics Research and Technology, School of Physics \& Astronomy, Cardiff University, Queens Buildings, \\
The Parade, Cardiff, CF24 3AA, UK\\
$^{22}$Department of Physics, University of Warwick, Gibbet Hill Road, Coventry CV4 7AL, UK\\
$^{23}$Birmingham Institute for Gravitational Wave Astronomy and School of Physics and Astronomy, University of Birmingham, Birmingham B15 2TT, UK \\
$^{24}$Isaac Newton Group (ING), Apt. de correos 321, E-38700, Santa Cruz de La Palma, Canary Islands, Spain\\
$^{25}$Astrophysics Research Centre, School of Mathematics and Physics, Queen’s University Belfast, Belfast BT7 1NN, UK
}
\date{Accepted XXX. Received YYY; in original form ZZZ}

\pubyear{2023}

\begin{document}
\label{firstpage}
\pagerange{\pageref{firstpage}--\pageref{lastpage}}
\maketitle

\begin{abstract}
We present an in-depth study of the late-time near-infrared plateau in Type Ia supernovae (SNe Ia), which occurs between 70-500 d. We double the existing sample of SNe Ia observed during the late-time near-infrared plateau with new observations taken with the Hubble Space Telescope, Gemini, New Technology Telescope, the 3.5m Calar Alto Telescope, and the Nordic Optical Telescope. Our sample consists of 24 nearby SNe Ia at redshift < 0.025. We are able to confirm that no plateau exists in the \textit{K$_s$} band for most normal SNe Ia. SNe Ia with broader optical light curves at peak tend to have a higher average brightness on the plateau in \textit{J} and \textit{H}, most likely due to a shallower decline in the preceding 100~d. SNe Ia that are more luminous at peak also show a steeper decline during the plateau phase in \textit{H}. We compare our data to state-of-the-art radiative transfer models of nebular SNe Ia in the near-infrared. We find good agreement with the sub-M$_{\rm ch}$ model that has reduced non-thermal ionisation rates, but no physical justification for reducing these rates has yet been proposed. An analysis of the spectral evolution during the plateau demonstrates that the ratio of \FeIIF\ to \FeIIIF\ contribution in a near-infrared filter determines the light curve evolution in said filter. We find that overluminous SNe decline slower during the plateau than expected from the trend seen for normal SNe Ia. 
\end{abstract}




\section{Introduction}

Although Type Ia supernovae (SNe Ia) are widely used as cosmic distance indicators \citep{Riess1998, Perlmutter1999}, there is still debate about their explosion mechanisms and the nature of their progenitors (see \citealt{Hillebrandt2013, Maoz2014, Ruiter2020, Jha2019} for comprehensive reviews). 
It is generally accepted that SNe Ia originate from the thermonuclear explosions of carbon-oxygen white dwarfs (CO WDs). The CO material burns to iron-group elements, and the radioactive decay of \nick\ $\rightarrow$ \cob\ ($t_{1/2}$ = 6 d) powers the early light curves of SNe Ia. Around 60 d post explosion the dominating radioactive decay chain shifts to the decay from \cob\ $\rightarrow$ \fe, with a longer half life of 78~d.

In the nebular phase (phase > 150 d), the outer layers of the SN ejecta become transparent and the inner regions of the ejecta become visible. Late-time spectroscopy can be used to search for hydrogen, which would point towards a single-degenerate scenario \citep{Hamuy2003, Mattila2005, Leonard2007, Shappee2013, Silverman2013, Lundqvist2013, Lundqvist2015, Maguire2016, Kollmeier2019, Graham2019, Prieto2020}. Late-time spectra can also be used to constrain the amount of stable nickel vs. unstable material, which can be compared to predictions from explosion models where a larger ratio points to burning at higher central densities, implying a larger progenitor mass \citep{Mazzali2015, Botyanszki2017, Maguire2018, Flors2020}.

The majority of studies at all epochs focus on the optical, because SNe Ia are brightest at these wavelengths and there are many optical instruments available. Studying the near-infrared (NIR, $\lambda>0.8$~µm) is more difficult, but is beneficial because SNe~Ia are better standard candles in this wavelength range \citep{Elias1981, Elias1985, Krisciunas2004, WoodVasey2008, Barone-Nugent2012, Johansson2021a, Jones2022, Galbany2022, Muller-Bravo2022a}. They are less impacted by extinction, with inferred distance estimate root mean square values dropping by 2--4$\sigma$ even before any corrections are applied to the light curves \citep{Avelino2019}.  

While most NIR studies focus on light curve standardisation around peak ($-$5--40 d), observations around 70--600 d are very important for understanding the evolution of the ejecta. Our understanding of the NIR evolution of SNe Ia at late times has evolved significantly during the past four decades. \cite{Axelrod1980} predicted an ``IR-catastrophe'' -- a shift from optical and NIR emission lines to fine-structure iron lines in the mid and far-IR that occurs at around 450 d due to an onset of thermal instability that causes a dramatic temperature change from $\sim$3000 K to $\sim$300 K. However, the resulting sharp decline in the optical and NIR light curves has never been observed in SNe Ia. \cite{Graur2020} found instead that SNe Ia reach a plateau in the $J$ and $H$ bands starting at $\sim$150 d and lasting for approximately a year. The presence of a plateau in the NIR was first predicted by \cite{Fransson1996}, where it was linked to the onset of the IR-catastrophe. 

\cite{Fransson1996} suggested the flattening of the \textit{J} band is due to the shift from emission of \FeIIIF\ at $\sim5000$ \AA\ to emission of \FeIIF\ at 1.257 µm and 1.644 µm, which is supported by the NIR evolution of SN 2014J presented by \cite{Diamond2018}. Updated spectral models by \cite{Fransson2015} showed that a redistribution of ultra-violet (UV) emissivity increases the flux in the optical and NIR, circumventing the ``IR-catastrophe''.

This flux-redistribution behaviour is reminiscent of the re-brightening seen in the NIR around 30 d past maximum, also called the secondary maximum. This feature is caused by sharp peaks in the emissivity of iron/cobalt gas at certain temperatures, which are near an ionisation edge \citep{Kasen2006}. The dependence of the emissivity on temperature explains why for subluminous and cooler 1991bg-like SNe Ia \citep{Filippenko1992, Leibundgut1993, Turatto1996, Taubenberger2017}, the secondary maximum is shifted to earlier phases, often causing it to blend with the primary maximum. The strongest peak in the NIR emissivity occurs at $\sim$7000 K and represents the ionisation edge between doubly ionised to singly ionised iron, when the ejecta becomes very efficient at redistributing flux from the UV to longer wavelengths, which leads to the re-brightening in the NIR during the secondary maximum. Another peak exists at the ionisation edge between singly ionised and neutral iron at $\sim$2500 K, which may coincide with the onset of the NIR plateau at $\sim$150~d. \cite{Diamond2018} presented NIR spectra of SN 2014J during the plateau phase, which demonstrated a decrease in the strength of \FeIIIF\ features in favour of \FeIIF\ features, but no \FeIF\ features. \cite{Sollerman2004} and \cite{Graur2020} agreed that the scattering of UV photons to longer wavelengths is the most likely cause of the NIR plateau. 

The end of the plateau at $\sim$500~d is not yet understood, although a tentative detection of \FeIF\ by \cite{Graur2020} suggested a third shift in the dominant ionisation state of iron. \cite{Tucker2022} also identified strengthening features in the optical after the end of the plateau that could be attributed to \FeIF. 

\cite{Graur2020} tentatively suggested that the plateau does not extend to the $K_s$-band, and that the plateau in the \textit{H} band is comprised of two distinct branches. No theoretical explanation for this bimodal behaviour was offered, although a correlation between the peak magnitude and the magnitude of the plateau would align well with the idea that the plateau is caused by a similar mechanism as the secondary maximum.

In this paper we extend the sample of SNe Ia with NIR photometry on the late-time plateau to 24 SNe Ia, and use the additional data to confirm the absence of the plateau in the \textit{K$_s$} band, and test whether the magnitudes in the \textit{H}-band plateau consist of two distinct branches as suggested by \cite{Graur2020}. In Section \ref{data}, we introduce the sample of nearby SNe Ia and describe the late-time NIR photometry and spectroscopy included in this paper. Fitting methods implemented in this work, as well as the radiative transfer models of SNe Ia in the nebular phase that we compare to our data are described in Section \ref{methods}. We present our results in Section \ref{results} and discuss their implications in a theoretical context in Section \ref{discussion}. Finally, we summarise and conclude in section \ref{conclusion}.

\section{Data}\label{data}

In Section \ref{sample}, we present the sample of nearby SNe Ia used in this paper. We describe the spectra and photometry included in this paper, which is a combination of data pulled from the literature and new data, in Section \ref{observations}.

\subsection{Sample of nearby SNe Ia}\label{sample}

Our sample consists of new data, as well as data that have previously been published, totalling 24 SNe Ia. Of these, 20 are classified as normal SNe Ia, two are classified as 91T-like (SNe 2000cx and 2021wuf), and two are classified as transitional objects (SNe 2004eo and 2012ht). The NIR photometry of six SNe Ia (SNe 2020ees, 2020uxz, 2021jad, 2021pit, 2021wuf, and 2021aefx), and XShooter spectra of two SNe Ia (SNe 2016hvl and 2017cbv) are presented for the first time in this paper. XShooter spectra of three SNe (SNe 2012cg, 2012ht, and 2013aa) published by \cite{Maguire2013}, and spectra of four SNe (SNe 2012fr, 2013cs, 2013ct, and 2013dy) published by \cite{Maguire2016} are also included.  We performed synthetic photometry on these spectra to extract NIR photometry as described in Section \ref{xshooter}. We include NIR photometry for ten SNe Ia presented by \cite{Graur2020} \citep[SNe 2000cx, 2001el, 2004eo, 2011fe, 2012ht, 2013dy, 2014J, 2017erp, 2018gv, and 2019np, which were originally published by][]{Krisciunas2003, Sollerman2004, Pastorello2007, Stritzinger2007, Sand2016, Shappee2017, Burns2018}. We note that the NIRI observations of SN 2020uxz were taken in \textit{K} rather than \textit{K$_s$}. The early observations of SN 2001el \citep{Krisciunas2003} are taken in a mixture of \textit{K} and \textit{K$_s$}, but here we only include the data taken in \textit{K$_s$}. The late-time observations from \cite{Stritzinger2007} are taken exclusively in \textit{K$_s$}. We never mix \textit{K} and \textit{K$_s$} data when performing fits, as will be discussed further in Section \ref{results}. An overview of the sample is presented in Table \ref{sample_overview}.

Observing the NIR plateau is difficult because SNe Ia are inherently fainter in the NIR compared to the optical and by 150 d they have faded by $\sim$6 magnitudes relative to peak. Consequently, all the SNe Ia in our sample are nearby, have $m_{\rm max}^{\rm B}$ $< 15$ mag (with the exception of SN 2016hvl which has $m_{\rm max}^{\rm B}$=15.4 mag), and are offset from their host galaxies to reduce host contamination. We note that the last criterion is one potential source of bias in our sample \citep{Wang2013}.

All distance moduli and uncertainties were taken either from the literature where available, or calculated from redshift-independent distances provided by the NASA/IPAC Extragalactic Database (NED).\footnote{\url{https://ned.ipac.caltech.edu/}} The data sources for the distance moduli are summarised in Table \ref{sample_overview}.

\subsection{Observations}\label{observations}

We report NIR photometry obtained using Wide-Field Camera 3 (WFC3) on the \textit{Hubble Space Telescope} (\textit{HST}), SOFI on the New Technology Telescope (NTT),\footnote{Data taken from the NTT were taken under the framework of the advanced Public ESO Spectroscopic Survey for Transient Objects \citep[ePESSTO+,][]{Smartt2015}.} FLAMINGOS-2 \citep[F2,][]{Eikenberry2008} at Gemini-South, NIRI at Gemini-North \citep{Hodapp2003}, and Omega2000 on the 3.5m Calar Alto Telescope \citep[CA,][]{Bailer-Jones2000, Baumeister2003, Kovacs2004} at the Centro Astronómico Hispano de Andalucía (CAHA), the Nordic Optical Telescope \citep[NOT,][]{Djupvik2010}, as well as spectra obtained with XShooter on the Very Large Telescope (VLT) at the Paranal Observatory \citep{Vernet2011}. The NIR photometry and spectroscopy are described in Sections \ref{photometry_description} and \ref{xshooter}, respectively. 

\subsubsection{NIR Photometry} \label{photometry_description}

Table \ref{phot_HST_table} shows the photometry of SNe 2020ees, 2020uxz, 2021jad, and 2021pit observed with WFC3 in the \textit{F125W} and \textit{F160W} filters, which can be approximated by the \textit{J} and \textit{H} bands, respectively (program ID's: GO-16497 and 16885, PI: Graur). We obtained photometry of SNe 2021jad, 2021pit, and 2021aefx with SOFI, which is a NIR spectrograph and imaging camera on the NTT (proposal ID's 1103.D-0328, 106.216C, and 108.220C, PI: Inserra). SN 2021pit was observed with F2 at Gemini-South (proposal ID: GS-2021B-FT-212, PI: Graur). \textit{H}-band photometry of SN 2020uxz and SN 2021wuf were obtained with NIRI at Gemini-North (proposal ID's: GN-2021A-FT-114 and GN-2022A-FT-210, PI: Graur and Deckers). Finally, SN 2020uxz was observed with Omega2000 (proposal ID: H20-3.5-002, PI: Galbany). SN 2020uxz was also observed with the NOT (proposal ID: 62-202, PI: Galbany). The photometry obtained with SOFI, F2, NIRI, Omega2000, and NOT are summarised in Table \ref{phot_table}.

All the photometry measurements were obtained using the package \textsc{autophot} \citep{Brennan2022}.\footnote{\url{https://github.com/Astro-Sean/autophot}} The data were calibrated using the 2MASS catalog in the Vega magnitude system. Since none of the sources are in very crowded fields, we implemented aperture photometry for the whole sample. As an additional test, point-spread function (PSF) photometry was performed where possible and compared to the aperture photometry. The aperture and PSF magnitudes were consistent within the uncertainties for all measurements. All sources are bright and far removed from their host galaxy and therefore template image subtraction was not required. Background surface fitting failed for a subset of the sample due to the noisy nature of the NIR images, so we reverted to local background fitting for the whole sample. 

No S-corrections were applied to the \textit{HST} photometry because no synchronous \textit{J}/\textit{F125W} or \textit{H}/\textit{F160W} data were available. We estimate the systematic offset between the filters by performing synthetic photometry on all the XShooter spectra in our sample for the \textit{J, H}, \textit{F125W} and \textit{F160W} bands. On average, we find that the \textit{F125W} photometry is 0.3 mag fainter than the \textit{J} band, and \textit{F160W} is 0.4 mag fainter than the \textit{H} band. We do not correct for these offsets but any \textit{HST}  photometry is highlighted in Fig. \ref{phot_plot} and the reader should note that these points are expected to be fainter than the corresponding ground-based filters.

Finally, all the data were corrected for Milky Way extinction using the dust map provided by \cite{Schlafly2011} and the Python module \textsc{dustmaps} \citep{Green2018}. The photometry was not corrected for host galaxy extinction because all the SNe are well separated from their host, and NIR photometry is minimally impacted by extinction. All the new NIR light curves, together with the NIR light curve data presented by \cite{Graur2020}, are shown in Fig. \ref{phot_plot}.

\begin{table*}
\begin{threeparttable}
 \caption{Overview of NIR photometry obtained with \textit{HST}. A machine readable version of this table is available in the online material.}\label{phot_HST_table}
\begin{tabular}{|l|c|r|l|r|l|}
\hline
     \textbf{SN} & \textbf{MJD} & \textbf{Phase}$^{a}$ & \textbf{Filter} & \textbf{Exposure} & \textbf{Magnitude}\\
      & [d] & [d] &  & \textbf{time} [s] & [mag] \\
     \hline
2020ees	&	58931.7	& 5.6  &	F125W&	46	 &		16.928(005) \\
2020ees	&	58931.7	& 5.6 &	F160W&	86	 &		17.056(005) \\
2020ees	&	58936.7	& 10.6 &	F125W&	46	 &		17.539(007) \\
2020ees	&	58936.7	& 10.6 &	F160W&	86	 &		17.486(006) \\
2020ees	&	59330.9	& 404.8 &	F125W&	1006	& 		23.066(037) \\
2020ees	&	59330.9	& 404.8 &	F160W&	1006	 &		22.822(042) \\
2020ees	&	59330.9	& 404.8 &	F350LP&	334	 	&	24.681(036) \\
2020ees	&	59439.3	& 513.2 &	F125W	&1006	& 		24.46(11) \\
2020ees	&	59439.3	& 513.2 &	F160W&	1006	& 		23.420(075) \\
2020ees	&	59439.3	& 513.2 &	F350LP&	334	 &		26.57(12) \\
2020uxz	&	59150.2	& 7.2 &	F125W&	18	 &		14.747(002) \\
2020uxz	&	59150.2	& 7.2 &	F160W&	29	& 		14.720(003) \\
2020uxz	&	59374.5	& 231.5&	F336W&	324	& 		20.713(022) \\
2020uxz	&	59374.5	& 231.5&	F350LP&	300	 &		19.554(003) \\
2020uxz	&	59374.5	& 231.5&	F125W&	203	 &		20.617(010) \\
2020uxz	&	59374.5	& 231.5&	F160W&	406	 &		20.265(013) \\
2020uxz	&	59485.2	& 342.2&	F336W&	 330     &	 		22.154(051) \\
2020uxz	&	59485.2	& 342.2&	F350LP&	   330   &	 		21.065(006) \\
2020uxz	&	59485.2	& 342.2&	F125W	&306	& 		20.698(011) \\
2020uxz	&	59485.2	& 342.2&	F160W	&306	& 		20.200(015) \\
2021jad	&	59623.0	& 294.3 &	F160W	&	356 & 	19.388(430) \\
2021jad	&	59819.0	& 490.3 &	F160W	&	356 & 	19.737(430)	 \\
2021jad	&	59973.5	& 644.8 &	F160W	&	356 & 	21.581(037)	 \\
2020pit	&	59630.0	& 245.5 &	F160W	&431 & 	18.594(050)	 \\
2020pit	&	59832.0	& 447.5 &	F160W	& 431	& 	19.428(050)	 \\
2020pit	&	59987.4	& 602.9 &	F160W	& 431	& 	21.941(037)	 \\

\hline
\end{tabular}
\begin{tablenotes}
    \footnotesize 
    \item [a] Phase is defined as the time since maximum light in the \textit{B} band (MJD - $t_0$).
\end{tablenotes}
\end{threeparttable}
\end{table*}

\begin{table*}
\begin{threeparttable}
 \caption{Overview of NIR photometry obtained with Gemini, the New Technology Telescope, and Calar Alto observatory. A full, machine readable version of this table is available in the online material.}\label{phot_table}
\begin{tabular}{|l|c|r|l|r|l|l|}
\hline
\textbf{SN} &  \textbf{MJD} & \textbf{Phase} & \textbf{Filter}& \textbf{Exposure}& \textbf{Magnitude}& \textbf{Instrument}\\
 & [d] & [d] &  & \textbf{time} [s] & [mag] &    \\
\hline
2020uxz & 	59178.0 & 	35.0 & 	J & 	60 & 	15.29(45) & 	NOTCAM \\ 
2020uxz & 	59178.0 & 	35.0 & 	H & 	60 & 	14.43(45) & 	NOTCAM \\ 
2020uxz & 	59184.9 & 	41.9 & 	J & 	600 & 	15.50(09) & 	Omega2000 \\ 
2020uxz & 	59184.9 & 	41.9 & 	H & 	900 & 	14.84(05) & 	Omega2000 \\ 
2020uxz & 	59434.6 & 	291.6 & 	H & 	111 & 	19.60(35) & 	NIRI \\ 
2020uxz & 	59434.6 & 	291.6 & 	J & 	111 & 	19.82(39) & 	NIRI \\ 
2020uxz & 	59434.6 & 	291.6 & 	K & 	298 & 	19.55(22) & 	NIRI \\ 
2021aefx & 	59563.3 & 	17.4 & 	J & 	32 & 	14.23(13) & 	SOFI \\ 
2021aefx & 	59563.3 & 	17.4 & 	H & 	32 & 	13.02(18) & 	SOFI \\ 
2021aefx & 	59591.0 & 	45.1 & 	J & 	24 & 	14.26(10) & 	SOFI \\  
2021aefx & 	59591.0 & 	45.1 & 	H & 	24 & 	13.48(19) & 	SOFI \\ 
2021aefx & 	59591.0 & 	45.1 & 	K$_s$ & 	60 & 	13.52(02) & 	SOFI \\ 
2021aefx & 	59612.1 & 	66.2 & 	J & 	32 & 	15.67(12) & 	SOFI \\ 
2021aefx & 	59612.1 & 	66.2 & 	H & 	32 & 	14.35(20) & 	SOFI \\ 
2021aefx & 	59640.1 & 	94.2 & 	J & 	32 & 	17.03(13) & 	SOFI \\ 
2021aefx & 	59640.1 & 	94.2 & 	H & 	32 & 	10.75(33) & 	SOFI \\ 
2021aefx & 	59640.1 & 	94.2 & 	K$_s$ & 	60 & 	11.30(08) & 	SOFI \\ 
2021aefx & 	59649.1 & 	103.2 & 	H & 	144 & 	16.46(48) & 	SOFI \\ 
2021aefx & 	59661.0 & 	115.1 & 	J & 	60 & 	>15.5$^{*}$ & 	SOFI \\ 
2021aefx & 	59661.0 & 	115.1 & 	H & 	144 & 	16.42(20) & 	SOFI \\ 
2021aefx & 	59661.1 & 	115.2 & 	K$_s$ & 	160 & 	15.97(19) & 	SOFI \\ 
2021aefx & 	59816.3 & 	270.4 & 	J & 	1080 & 	18.89(24) & 	SOFI \\ 
2021aefx & 	59816.3 &   270.4 & 	H & 	1440	 & 18.31(18) & 	SOFI \\ 
2021aefx & 	59816.3 & 	270.4 & 	K$_s$ & 	1440 & 	18.33(21) & 	SOFI \\ 
2021jad & 	59492.3 & 	163.6 & 	J & 	1080 & 	18.81(16) & 	SOFI \\ 
2021jad & 	59492.3 & 	163.6 & 	H & 	1440 & 	17.81(11) & 	SOFI \\ 
2021jad & 	59513.3 & 	184.6 & 	J & 	1080 & 	19.16(15) & 	SOFI \\ 
\hline
\end{tabular}
\begin{tablenotes}
    \footnotesize 
    \item [*] We found a very large uncertainty on the magnitude (18.5 $\pm$ 3.0 mag) for one \textit{J}-band image of SN 2021aefx at MJD = 59661.0~d. This is likely because the source was faint and the exposure time (60 s) was not sufficient. The next data point at MJD = 59816.3 d has a similar magnitude (18.89 $\pm$ 0.24 mag) but was exposed for 1080 s and has a significantly smaller uncertainty. We quote this data point as an upper limit at 15.5 mag.
\end{tablenotes}
\end{threeparttable}
\end{table*}

\begin{table*}
\caption{Overview of synthetic NIR Photometry obtained with XShooter.}\label{xshooter_table}
\begin{tabular}{|l|r|r|c|c|c|c|l|}
\hline
     \textbf{SN} & \textbf{MJD} & \textbf{Phase} & \textbf{\textit{J}}& \textbf{\textit{H}}& \textbf{\textit{K$_s$}} & \textbf{Flux calibration} & \textbf{Source of flux} \\
     & [d] & [d] & [mag] & [mag]& [mag] &  \textbf{uncertainty} [mag] &  \textbf{calibration}\\
     \hline
2012cg & 56420.0 &  337.8  &  19.09 & 18.29  & 20.48  &  0.2 & SDSS  \\
2012fr & 56600.0 & 358.1  &18.93   &  18.03 &  17.81 &  0.5 &  XShooter zeropoint \\
2012ht & 56728.0 & 432.9  & 20.60  &  20.01 & -  &  0.5 &  SDSS \\
2013aa & 56704.0 &  361.5  & 19.30  & 18.53  & 18.94  &  0.8 &  XShooter zeropoint  \\
2013aa & 56768.0 & 425.5  & 19.29  &  18.56 & 19.87  &  0.5 &  XShooter zeropoint  \\
2013cs &  56741.0 & 303.8 &  21.41 & 20.55  &  - & 0.5  &  XShooter zeropoint  \\
2013ct & 56615.0 & 198.9  &  18.40 & 17.38  & 17.23  & 0.2  & SDSS   \\
2016hvl & 58072.3 & 361.4   & 22.74  &  21.60 &  20.06 & 0.2  &  PS1 \\
2017erp & 58225.3 & 290.7  &  19.68 &  18.83 &  - & 0.2  & LCO Photometry  \\
2017erp & 58282.2 &  347.6 &  20.05 &  19.12 &  - &  0.1 &  LCO Photometry  \\
2017erp & 58308.1 & 373.5  &  20.03 &  19.23 & -  &  0.1 & LCO Photometry   \\
    \hline
\end{tabular}
\end{table*}

\subsubsection{Spectroscopy} \label{xshooter}

We include 12 mid-resolution spectra of eight SNe~Ia obtained using XShooter. Eight of these spectra were previously presented by \cite{Maguire2013, Maguire2016}. Four spectra are published here for the first time and were reduced using the same method described by \cite{Maguire2016}. Due to the relatively high spectral resolution of XShooter ($\sim$35 km s$^{-1}$), host galaxy features were easily identified and removed in the reduction process \citep{Maguire2016}. We do not expect to see any contribution from a potential companion star since the remnant models presented in \citep{Pan2012} predict that the contribution will be very faint relative to the SN at these phases. The spectral response of XShooter is relatively stable with a relative flux uncertainty across the spectrum of 5 per cent \citep{Vernet2011}. The three arms of the XShooter spectrograph were firstly combined using their overlap wavelength regions with small scalings in their flux levels. Next, the spectra were flux calibrated using photometry from Las Cumbres Observatory Global Telescope Network \citep[LCO;][]{Brown2013} if possible, or alternatively, using photometry performed on stars in the Sloan Digital Sky Survey (SDSS) or Pan-STARRS1 (PS1) field of the acquisition image. As a last resort, zeropoints were taken from XShooter. 

The XShooter acquisition image of each SN was used to estimate the magnitude of the SN and comparison stars in the field of the SN. The spectra of SNe 2012cg, 2012ht, and 2013ct were calibrated by comparison of the companion stars to catalogue magnitudes from the SDSS Data Release 10 \citep{Ahn2014}, and the spectrum of SN 2016hvl was calibrated by comparison to the PS1 Data Release 2 catalog \citep{Flewelling2020}. The spectrum of SN 2017erp was calibrated to LCO photometry of the SN itself taken at similar phases to the spectral observation. The spectra of SNe 2012fr, 2013aa, and 2013cs were calibrated using the XShooter zero-points because no coeval SN photometry nor catalogue magnitudes from SDSS or PS1 were available. In these cases, the tabulated XShooter zero-point was used, resulting in a larger uncertainty. The uncertainty was estimated by comparing the magnitudes obtained using the zero-point method for SNe that also had catalogue measurements. These were found to be $<$0.5 mag, which we set as the conservative uncertainty of the magnitudes estimated using the zero-point method. The different sources for flux calibration result in a large range of uncertainties. A summary of the flux calibrations is presented in Table \ref{xshooter_table}.

\begin{figure*}
    \centering
    \includegraphics[width=17cm]{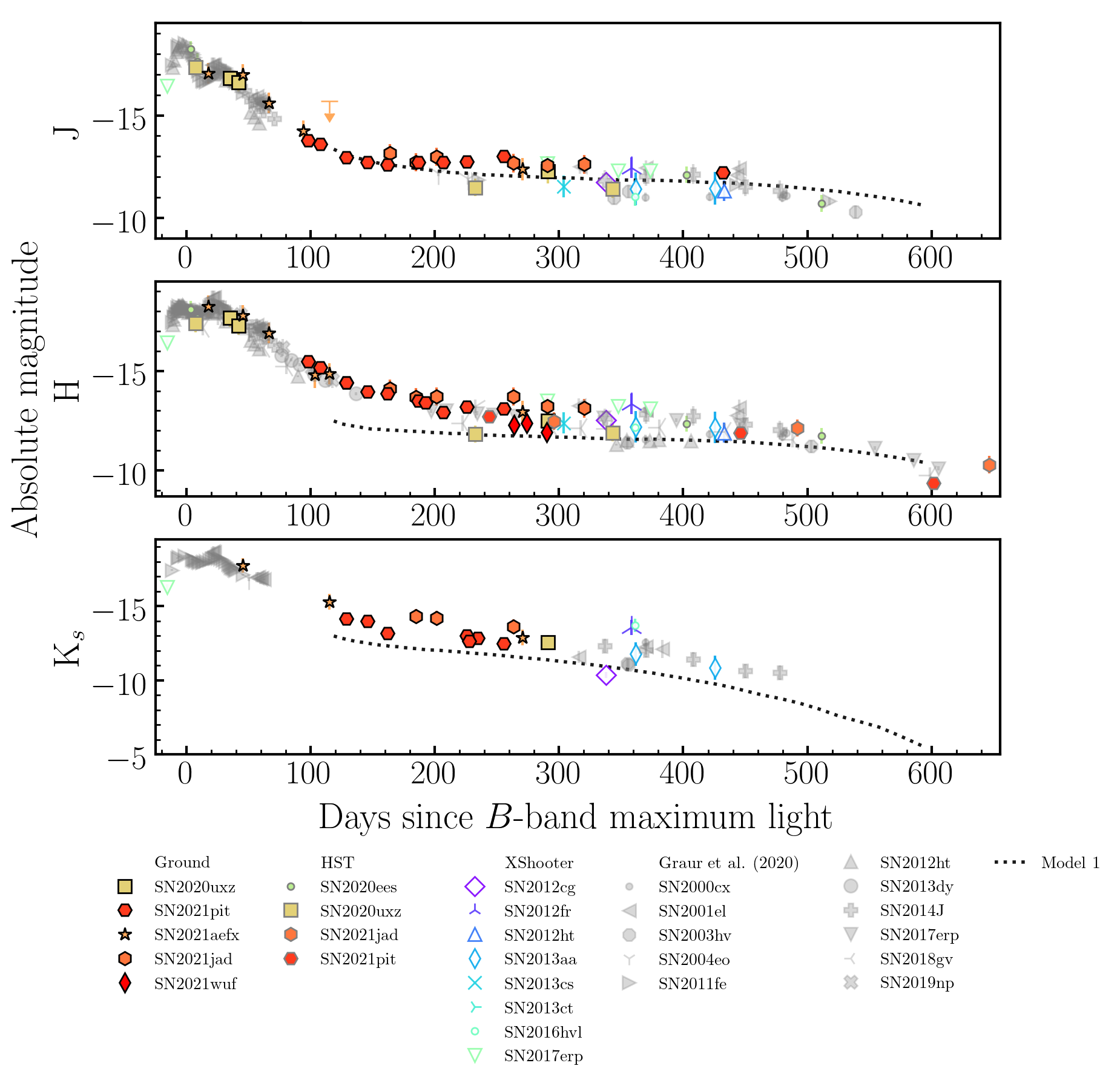}
    \caption{We present an overview of the \textit{J}-, \textit{H}-, and \textit{K$_s$}-band light curves in absolute magnitude for all the SNe Ia in our sample. We note that the NIRI photometry of SN 2020uxz is taken in \textit{K} rather than \textit{K$_s$}. We also include a comparison to the sub-\mch\ model with 8x heatboost that best matched SN 2013ct from \protect\cite{Shingles2022}. The model is scaled to our photometry in the \textit{J} band.}
    \label{phot_plot}
\end{figure*}

XShooter spectra extend from 5000 \AA\ to 25000 \AA, but in some cases the spectrum is very noisy at the far red end. We excluded spectra with spurious flux values at the red edge of the detector by visual inspection. We used \textsc{sncosmo} \citep{Barbary2022} to integrate across the \textit{J}, \textit{H}, and \textit{K$_s$} 2MASS bandpasses to obtain synthetic photometry (see Table \ref{xshooter_table} and Fig. \ref{phot_plot}).

\section{Methods}\label{methods}

In Section \ref{fitting_nir_plateau}, we describe how we fit the NIR data to determine if there is a plateau, and how we derive the average magnitude and decline rate of the plateau. In Section \ref{salt_fits}, we describe SALT3 fits performed on the optical light curves around peak. We describe the radiative transfer models of SNe Ia in the nebular phase, which were first presented by \cite{Shingles2022}, in Section \ref{model_description}.

\subsection{Fitting the NIR data}\label{fitting_nir_plateau}

To determine whether a light curve displays a plateau, and if so, when the transition onto the plateau occurs, we performed one- and two-component fits to the light curves between 30 -- 500 d using a Markov Chain Monte Carlo (MCMC) using the package \textsc{emcee} \citep{Foreman-Mackey2013}. If a two-component fit is preferred over a one-component fit, we classify the light curve as having a plateau. We opted to use MCMC to perform these fits to obtain robust estimates of the uncertainties on each parameter. 

For the one-component fit, we fit the following equation to $m^x(t)$, the magnitude in filter $x$ at time $t$:
\begin{equation}
    m^x(t) = s_2 t + b_2
\end{equation}
where $s_2$ is the slope and $b_2$ is the y-intercept. For the two-component fit, we implemented the same method as that used by \cite{Anderson2014} for characterising the light curves of SNe II. The two-component fit is described by the following piece-wise function:
\begin{equation}
m^x(t)=\begin{cases}
          s_1 t + b_1 \quad &\text{if} \,\ t<=t_0^x \\
          s_2 t + b_2 \quad &\text{if} \,\ t>t_0^x \\
     \end{cases}
\end{equation}
Where $s_2$ and $b_2$ are the same as for the one-component fit, and $s_1$ and $b_1$ are the slope and y-intercept of the function prior to the transition onto the plateau. The time of the onset of the plateau in filter $x$, $t_0^x$ is defined as: 
\begin{equation}
    t_0^x = \frac{b_2 - b_1}{s_1 - s_2}
\end{equation}
to ensure that the two linear components intersect at $t_0^x$. 

We ran an MCMC using 10 walkers for 10,000 iterations and uninformative priors. To avoid biasing the estimates of the slope during the plateau, we exclusively used data taken in \textit{J}/\textit{H} or \textit{F125W}/\textit{F160W}. We required at least four data points to perform the two-component fit since we are fitting for four parameters, and we required at least one data point at < 150 d and one at > 150 d to ensure we are sampling the phase ranges at either side of the expected transition onto the plateau. Only SNe 2001el, 2011fe, 2012ht, 2014J, 2018gv, 2021pit, and 2021aefx had sufficient data coverage to perform both one- and two-component fits across the range 30 -- 500~d. For the rest of the sample, there is not enough data to determine the plateau onset and we only performed one-component fits between 150 -- 500 d to find a single slope and y-intercept ($s_2$, $b_2$). For the objects where only the one-component fit was possible, we limited the phase range to 150 -- 500 d because it is unclear whether these SNe display a plateau phase, and we want to ensure we do not include data before the transition onto the plateau. At least two data points were required per band per SN to perform the one-component fit. 

On short timescales, the photometric uncertainty dominates over the temporal evolution, which results in highly uncertain estimates of the slope. We therefore required at least two data points to separated by at least 25 d, which reduced the number of SNe Ia with a measurement of the decline rate to 14. The minimum spacing of 25 d was determined by comparing the expected evolution with the expectation fluctuation within uncertainties. The mean uncertainty on the magnitude across our sample is 0.3 mag. The decline rate in the \textit{K$_s$} band measured across the whole sample is 1.2 $\pm$ 0.2 mag~/~100~d, meaning that a change of $\sim$0.3 mag would be expected to occur across approximately 25~d. 

We used reduced-$\chi^2$ ($\chi^2_{\rm red}$) to describe the quality of a fit and we used the Akaike Information Criterion \citep[AIC;][]{Burnham2004} to determine the rank of the one- and two-component fits. AIC penalizes extra degrees of freedom to avoid over-fitting the data, and is defined as follows:
\begin{equation}
AIC = -2 \ln(L) + 2k
\end{equation}
where $L$ is the likelihood and $k$ is the number of free parameters. For comparing the one- and two-component models, we have $k=2, 4$, respectively. Therefore, the two-component model is penalised for its two additional degrees of freedom by four AIC units. If the one- and two-component models differ by more than 2 AIC units, the model with a lower score was deemed the better fit. The best matching value is taken from the 50th percentile, and the uncertainties were taken from the 16th and 84th percentiles of the marginalised distribution.

We used the best matching fits to determine the properties of the plateau for each SN. The decline rate during the plateau phase is taken as the slope ($s_2$), which we quote in units of mag~/~100~d. The average magnitude is calculated from $b_2$ and $s_2$ of the best matching fit.

To minimise the impact from poorly sampled light curves, we also performed the one- and two-component fits for the full combined sample in absolute magnitude for each filter. Since there is minimal intrinsic scatter in the NIR, this should give a good estimate of the average decline rate for each filter.

\subsection{SALT3 light curve fits}\label{salt_fits}
In order to determine the general properties of each SN, we fitted optical light curves taken from the literature with the package \textsc{sncosmo}, using the SALT3 model \citep{Kenworthy2021}. The sources of the optical light curves are listed in Table \ref{sample_overview}. For SNe with no published optical data, we used preliminary photometry from LCO provided by the Global Supernova Project (GSP). We excluded any UV or NIR data because SALT3 is not well trained at those wavelengths, and we restricted the data to between $-$10 d to +40~d. The SALT3 parameters derived from these fits ($x_1$, a metric of the light curve stretch, and $c$, a measure of the colour at peak) are presented in Table \ref{sample_overview}. There was no optical light curve available for SN 2013ct so we were not able to derive SALT3 parameters.

\subsection{Comparison to radiative transfer models}\label{model_description}

We compared our sample to the sub-\mch\ SN Ia models of \cite{Shingles2022}. These models use the \cite{Shen2018} model of a detonation of a 1 $\msun$ WD and evolve the post-explosion composition using the radiative transfer code \textsc{artis}. Earlier models by \cite{Fransson1996} predicted a strong decline in the optical as flux is redistributed to the NIR, which was not matched by observations. The improved treatment of non-local scattering and fluorescence by \cite{Fransson2015} alleviated some of the discrepancies between the models and the observations, but no light curves were published for direct comparison. The models of \cite{Shingles2022} use a modified treatment of non-thermal energy deposition in which the the energy loss to free electrons is artificially boosted as a way to lower the ionisation state. With this modification, the models are able to reconcile the strength of the \FeIIF\ features, which are generally under-produced by sub-\mch\ models. Others have suggested that clumping of the ejecta is required to reduce the ionization state \citep{Wilk2018}. The sub-\mch\ model with a plasma loss rate increased eight-fold (model 1) is best able to reproduce the nebular NIR spectrum of the normal SN 2013ct \citep[see fig. 6 in][]{Shingles2022}. In this work, we present a time-extended version of the sub-M$_{\rm ch}$-heatboost8 model. For further details of the the model, we refer the reader to \cite{Shingles2022}. We also include the other three sub-M$_{\rm ch}$ models presented by \cite{Shingles2022} (sub-M$_{\rm ch}$-heatboost$\times$4, sub-M$_{\rm ch}$, sub-M$_{\rm ch}$-AxelrodNT), referred to from hereon as models 2, 3, and 4, respectively. However, because SN 2013ct is a normal SN Ia and is included in our sample, we focus on the best-matching model to its nebular spectrum (model 1). In Fig. \ref{phot_plot} the model is scaled to the \textit{J}-band photometry from our sample.

\section{Results} \label{results}

We constrain the onset of the plateau for a sub-set of SNe Ia in Section \ref{onset_fits}. In Sections \ref{decline_results} and \ref{ave_mag_text}, we present the decline rates and average magnitudes during the plateau of the SNe Ia in our sample. We compare the NIR plateau properties to SN properties at peak in Section \ref{correlations}.

\subsection{Properties of the NIR plateau} \label{photometric_properties}

We analyse the photometry presented in Section \ref{data} and shown in Fig. \ref{phot_plot} using three metrics: the onset of the plateau, the decline rate, and the average magnitude during the plateau. These metrics are derived from the fits either to each individual SN or to the sample as a whole in each filter, as described in Section \ref{fitting_nir_plateau}.

\begin{table*}
\begin{threeparttable}
\caption{Summary of parameters for the fits performed for SNe Ia with data available before and after the transition onto the plateau. We perform both a one- and two-component fit to each light curve. SNe Ia with a maximum separation of less than 25 d between data points are not included (see Section \ref{decline_results}).}\label{fit_parameters}
\begin{tabular}{|l|l|c|c|c|r|r|r|}
\hline
     SN & Filter & $s_1$ & $s_2$ & $t_0$ & $\chi^2_{\rm red}$ & $\chi^2_{\rm red}$ & $\Delta$AIC \\
      & & [mag/100 d] & [mag/100 d]  & [d] & one-comp. & two-comp. &   \\
      (1) & (2) & (3) & (4) & (5) & (6) & (7) & (8)  \\
     \hline
Full sample                     & J             &   5.2$^{+0.9}_{-0.5}$       &   0.4 $\pm$ 0.1              &    120 $\pm$ 10   &  29 &  14 &  1137  \\  
Full sample                     & H             &    3.5$^{+0.3}_{-0.3}$      &   0.5 $\pm$ 0.1               &    140 $\pm$ 10   &  63 & 29 &  3832  \\     
Full sample                     & K$_s$             &    -      &      1.2 $\pm$ 0.4            &   -    & 33 & 42 &   -407 \\     
SN 2000cx                     & J            &         -         & -0.1 $\pm$ 0.1              &         -     &   0.1 & - &  -    \\
SN 2000cx                     & H           &          -         & -0.2 $\pm$ 0.1             &         -       &  0.2 & - &  -  \\
SN 2001el                     & J            &        7.9$^{+5.9}_{-4.6}$        & 0.3 $\pm$ 0.2          & 90$^{+40}_{-50}$         &   0.7 & 0.08 & 2          \\
SN 2001el                     & H            &        8.3$^{+7.1}_{-4.9}$        & 0.3 $\pm$ 0.3              & 100$^{+40}_{-50}$          &    1.2 & 0.2 &  4       \\
SN 2001el                     & K$_s$            &         -         & 1.7 $\pm$ 0.1             &          -     &  1.2 & 1.7 &  -4         \\
SN 2003hv                     & H            &         -         & 0.1 $\pm$ 0.3                &           -    &      0.03 & - &       \\
SN 2011fe                     & J             &        8.8 $\pm$ 0.5    & 0.1 $\pm$ 0.5               &     90 $\pm$ 3      & 80 & 0.1 & 790   \\
SN 2012ht                     & J             &     7.9 $\pm$ 0.2     & 0.3 $\pm$ 0.3               &               90 $\pm$ 10    &   274 & 0.1 & 1093  \\
SN 2012ht$^*$                    & H             &       4.9$^{+0.3}_{-0.2}$    & 0.1 $\pm$ 0.1   &         150 $\pm$ 10   &    230 & 4 &  1113  \\
SN 2013aa                     & J             &        -         & 0.3 $\pm$ 1.2              &            -          & 0.1 & - &  -  \\
SN 2013aa                     & H             &        -         & 0.3 $\pm$ 1.2               &               -      &   0.01 & - &  -  \\
SN 2013aa                     & K$_s$             &        -         & 0.8$^{+1.1}_{-1.4}$                &             -        &  0.2 & - &  -  \\
SN 2014J                      & J             &      5.7 $\pm$ 0.3      & 0.4 $\pm$ 0.3                 &              100 $\pm$ 20        &  17 & 1 & 213\\
SN 2014J                      & H             &      4.1 $\pm$ 0.4       & 0.5 $\pm$ 0.3                &             130 $\pm$ 40    &    8 & 2 & 81     \\
SN 2014J                      & K$_s$             &        -         & 1.3$^{+0.1}_{-0.2}$              &               -         & 0.06 & - & - \\
SN 2017erp                    & J             &        -         & 0.5 $\pm$ 0.4                &                 -     & 0.1 & - &  -  \\
SN 2017erp                    & H             &         -        & 0.5 $\pm$ 0.4                &                -      &  0.01 & - &  -  \\
SN 2018gv                     & F160W             &       5.4$^{+1.8}_{-1.7}$      & 0.3 $\pm$ 0.2              &             130$^{+20}_{-30}$       &  4 & 0.1 & 31  \\
SN 2020uxz                    & J             &         -        & 0.1$^{+0.9}_{-0.3}$                &            -       &    0.001 & - &  -  \\
SN 2020uxz                    & H             &         -        & 0.0$^{+0.9}_{-0.4}$             &               -     &     0.001 & - &  -  \\
SN 2021jad                    & J             &          -       & 0.3 $\pm$ 0.1                &              -    &    0.1 & - &  -    \\
SN 2021jad                    & H             &         -        & 0.5 $\pm$ 0.1                  &             -     &    0.1 & - &  -  \\
SN 2021jad                    & K$_s$             &         -        & 0.9 $\pm$ 0.4               &               -    &    0.01 & - &  -    \\
SN 2021pit                    & J             &       3.1$^{+1.7}_{-0.7}$      & 0.2 $\pm$ 0.1                &     130$^{+30}_{-70}$   &  6 & 2 & 40  \\
SN 2021pit$^*$                    & H             &       3.5$^{+3.5}_{-0.6}$     & 0.6 $\pm$ 0.1                &       160$^{+30}_{-40}$     &   10 & 4 &  50\\
SN 2021pit                    & K$_s$             &          -       & 1.3$^{+0.2}_{-0.3}$              &                  -    & 4  & 6 & -5  \\
SN 2021aefx                    & J             &         6.0$^{+2.9}_{-2.5}$      &  0.9 $\pm$ 0.5              &         & 2.5  & 0.3 & 3 \\
SN 2021aefx                    & H             &         5.2$^{+2.5}_{-1.8}$      & 1.0$^{+0.5}_{-0.6}$                &                  100$^{+60}_{-50}$  &  2 & 0.2 & 3  \\
SN 2021aefx                    & K$_s$             &     -         &       1.6$^{+0.3}_{-0.5}$         &     -               &  0.01 & - & -  \\
SN 2021wuf                    & H             &          -       & 1.1$^{+1.1}_{-1.0}$               &                  -     &  0.1 & - &  - \\
\hline
\end{tabular}
\begin{tablenotes}
 \footnotesize 
        \item \textsc{note}-- Columns (3) \& (4): $s_1$ and $s_2$ are the slopes prior to and during the plateau. We also include $s_2$ for SNe Ia that exclusively have data during the plateau, for which we performed only one-component fits. Column (5): $t_0$ is the phase at which the SN transitions onto the plateau. Column (6) \& (7): $\chi^2_{\rm red}$ values of the one- and two-component fits, describing the quality of the best-matching fit. Columns (8): The difference between the AIC values for the one- and two-component fits ($\Delta$AIC = AIC$_{\rm one-comp.}$ $-$ AIC$_{\rm two-comp.}$). If the AIC values of two models differ by more than 2 units, the model with the lower AIC value is deemed significantly better, and the parameters for that model are quoted. 
        \item[*] The pre- and post-transition data have contributions from both space and ground based telescopes.
\end{tablenotes}
\end{threeparttable}
\end{table*}

 \subsubsection{Constraining the onset of the plateau}\label{onset_fits}

SNe 2001el, 2011fe, 2012ht, 2014J, 2018gv, 2021pit, and 2021aefx have data before and during the plateau, enabling us to constrain the phase of the onset of the plateau. The onset of the plateau is calculated by fitting a two-component linear fit to the light curves, as described in Section \ref{fitting_nir_plateau} and shown in Fig. \ref{slope_two_component}. We also fit each SN with a one-component fit and compare the result to the two-component fit using the AIC. If the one- and two-component fits differ by more than two AIC units, the model with the lower AIC is deemed significantly better. 

SN 2021pit, which is the best sampled SN along the transition onto the plateau, is best fit with two components in \textit{J} and \textit{H}, and yields $t_0^J$ = 130$^{+30}_{-70}$~d and $t_0^H$ = 160$^{+30}_{-40}$~d. The \textit{K$_s$} band is best fit with a single-component decline ($\Delta$AIC = 5). The uncertainty on the time of the onset of the plateau is large and we are unable to constrain the onset to the order of a few days, likely due to the gradual nature of the transition. 

The results from the fits to SNe 2011fe, 2012ht, 2014J, 2018gv, and 2021aefx are summarised in Table \ref{fit_parameters}. All \textit{J}- and \textit{H}-band light curves are best matched by a two-component model. Only SNe 2001el and 2021pit have sufficient \textit{K$_s$}-band data to perform one- and two-component fits, and they were both best matched by a one-component model suggesting no plateau exists in this band. 

We find that the $t_0^J$ and $t_0^H$ values for the other SNe are consistent with those derived for SN 2021pit. We note that SN 2012ht has pre-transition data in \textit{F160W} whereas the post-transition data is taken with \textit{H}, therefore the parameters derived describing the transition onto the plateau should be treated with caution. Similarly, SN 2021pit has post-transition data from \textit{HST} in \textit{F160W}. However, repeating the fit excluding the \textit{HST} data produces consistent results ($t_0^H$ = 140~$\pm$~30 d). 

We calculate the weighted mean of $t_0$ for all measurements across one filter, taking into account the uncertainties, and find $t_0^J$ = 90~$\pm$~20~d and $t_0^H$ = 130~$\pm$~20~d, where the uncertainties are quoted as the standard deviations. This implies $t_0^J$ and $t_0^H$ are consistent. This disagrees with the trend for the secondary maximum, where the second peak occurs in \textit{H} before it occurs in \textit{J} \citep{Kasen2006, Dhawan2015}. However, a larger, better sampled collection of SNe is required to reduce the uncertainties on the time of the transition and test if the time of transition is truly consistent between the \textit{J} and \textit{H} bands. 

To increase the sample size, we repeat the same one- and two-component fits for the full combined sample (Fig. \ref{slope_two_component}). We find that the \textit{J}- and \textit{H}-band data are best fit with two components ($\Delta$AIC = 1137 and 3832), with $t_0^J$ = 120~$\pm$~10~d and $t_0^H$ = 140~$\pm$~10~d. The \textit{K$_s$} band is best fit with a single, constantly declining component ($\Delta$AIC = 580).

\subsubsection{Decline rate during the plateau}\label{decline_results}

\begin{figure}
\captionsetup[subfigure]{labelformat=empty}
 \centering
     \begin{subfigure}[b]{0.45\textwidth}
         \centering
         \includegraphics[width=8cm]{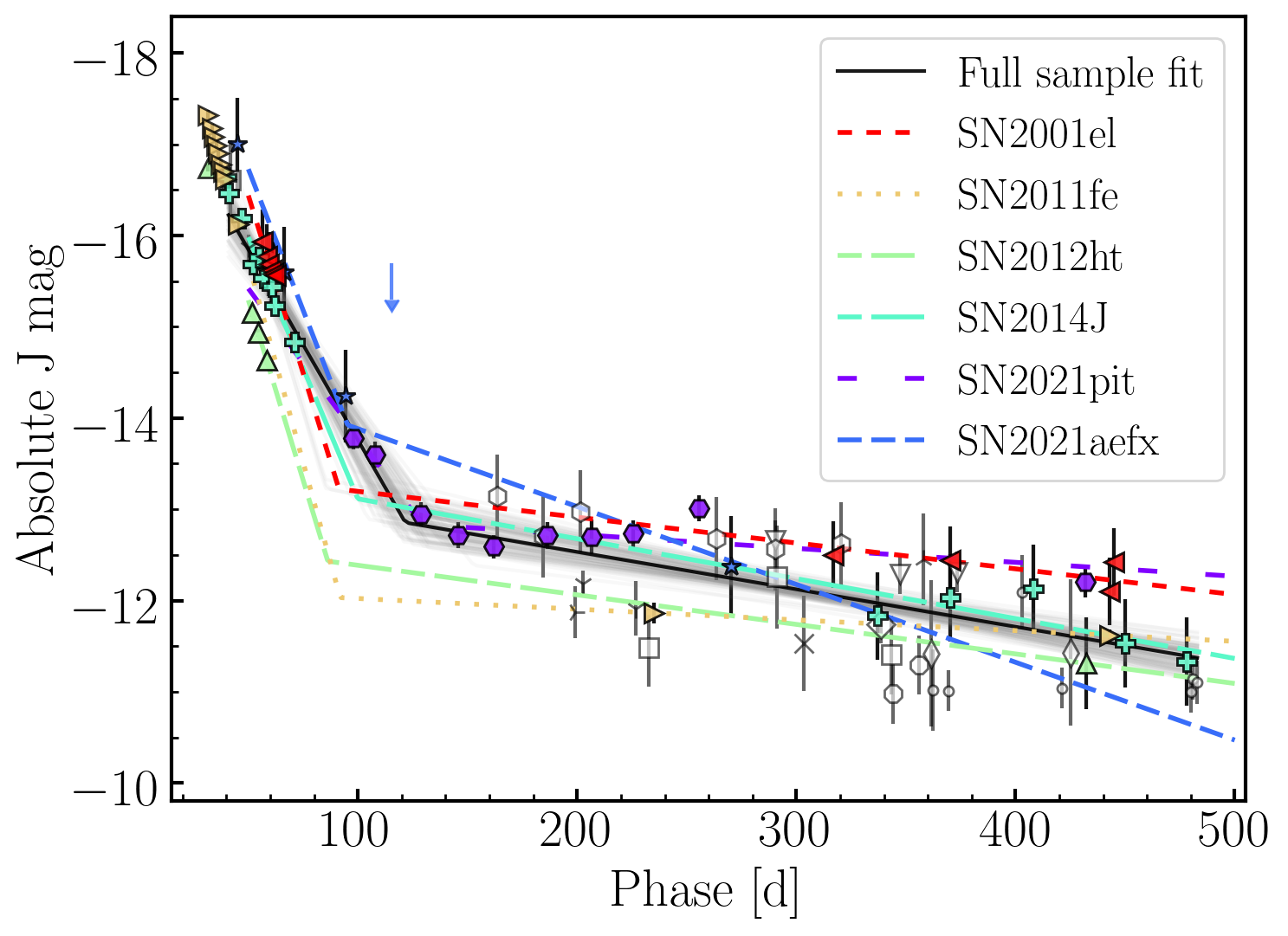}
         \caption{}
     \end{subfigure}%
        \\
     \begin{subfigure}[b]{0.45\textwidth}
         \centering
         \includegraphics[width=8cm]{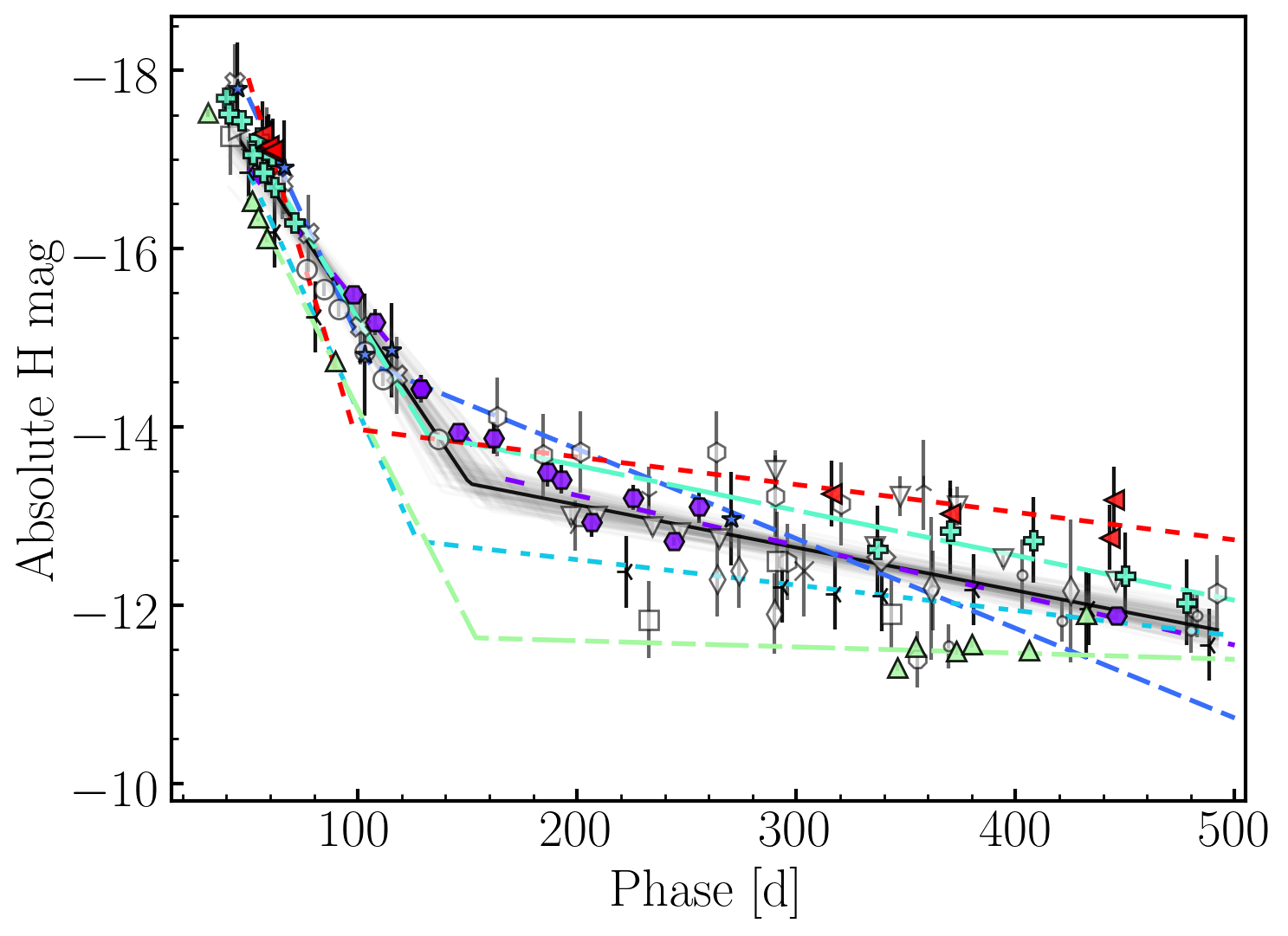}
         \caption{}
     \end{subfigure}%
     \\
     \begin{subfigure}[b]{0.45\textwidth}
         \centering
         \includegraphics[width=8.cm]{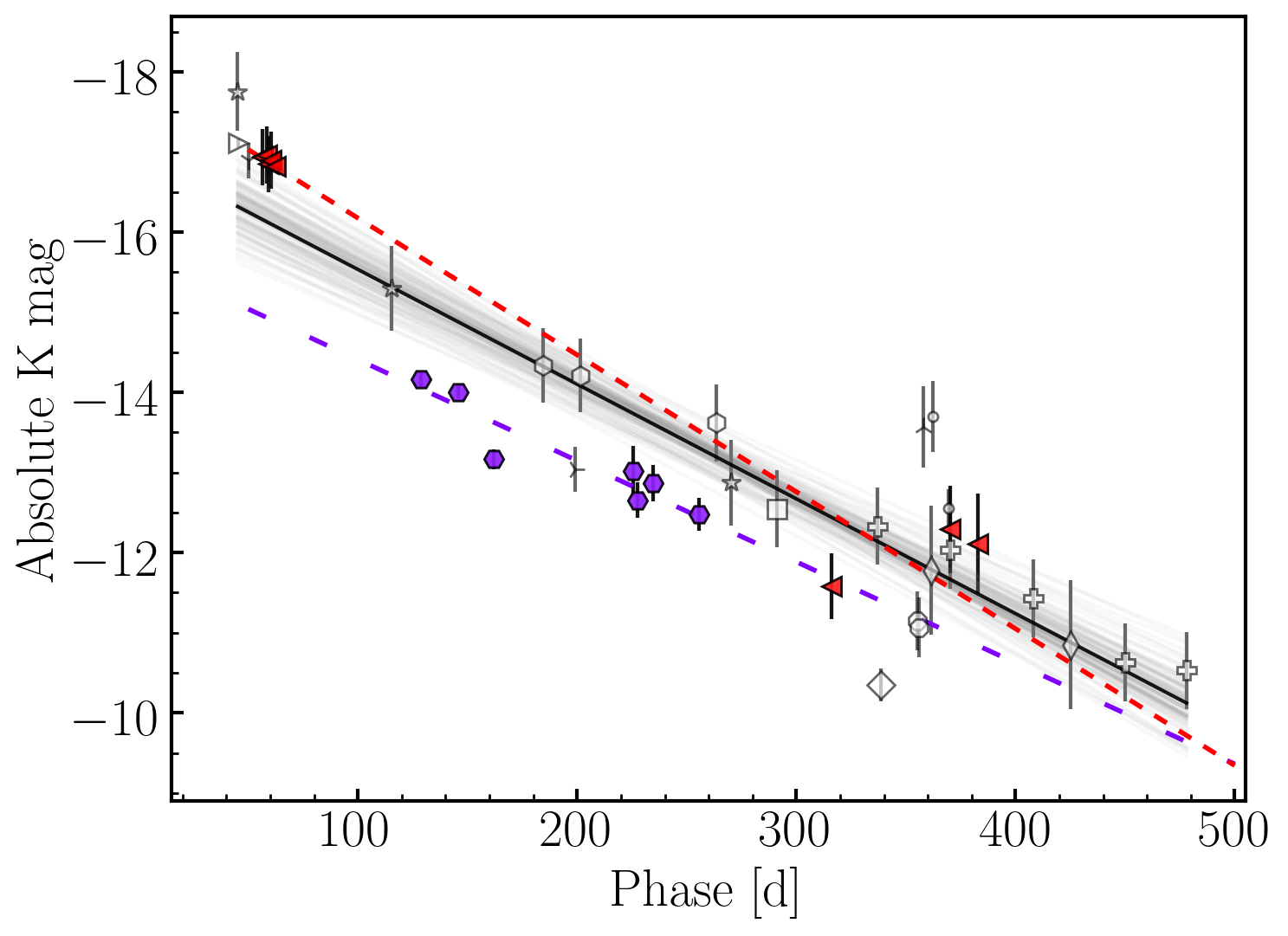}
         \caption{}
     \end{subfigure}
        \caption{The observed light curves and two-component MCMC fits for the \textit{J} (top), \textit{H} (middle), and \textit{K$_s$} (bottom) filters. The best fits (one- or two-component) to the NIR light curves of the whole sample are shown as black solid lines, with the various MCMC iterations shown as faded grey lines. Both the \textit{J} and \textit{H} bands have a non-zero decline rate but are consistent with zero within 2$\sigma$, and are best fit with two components. The decline rate in the \textit{K$_s$} band is inconsistent with zero at a >6$\sigma$ confidence level, and is best matched by a one-component fit. We also show the best matching fits to the SNe Ia with sufficient data (markers are the same as in Fig. \ref{phot_plot}).}
        \label{slope_two_component}
\end{figure}

In Fig. \ref{slope_two_component} we show two-component linear fits, fitted to all the SNe Ia simultaneously in absolute magnitude in each filter. By fitting the full combined sample, the influence from a single, potentially poorly sampled SN, is minimised. In the \textit{J} and \textit{H} bands, the decline rates of the second component (during the plateau) are $s_2^J$ = 0.4 $\pm$ 0.1 and $s_2^H$ = 0.5 $\pm$ 0.1 mag~/~100~d, respectively. These are inconsistent with zero at a >3$\sigma$ confidence level, meaning that the decline does not cease completely during the plateau. However, by comparing to the decline rate prior to the plateau ($s_1^J$ = 4.3 $\pm$ 0.6 and $s_1^H$ = 3.6$^{+0.4}_{-0.3}$ mag~/~100~d) it is clear that the decline slows significantly. The decline rate in \textit{K$_s$} is inconsistent with zero at a >6$\sigma$ confidence level, since this band is best fit by a single component with a continuous decline ($s_2^K$ = 1.2 $\pm$ 0.2 mag~/~100~d).

\begin{figure}
    \centering
    \includegraphics[width=8.4cm]{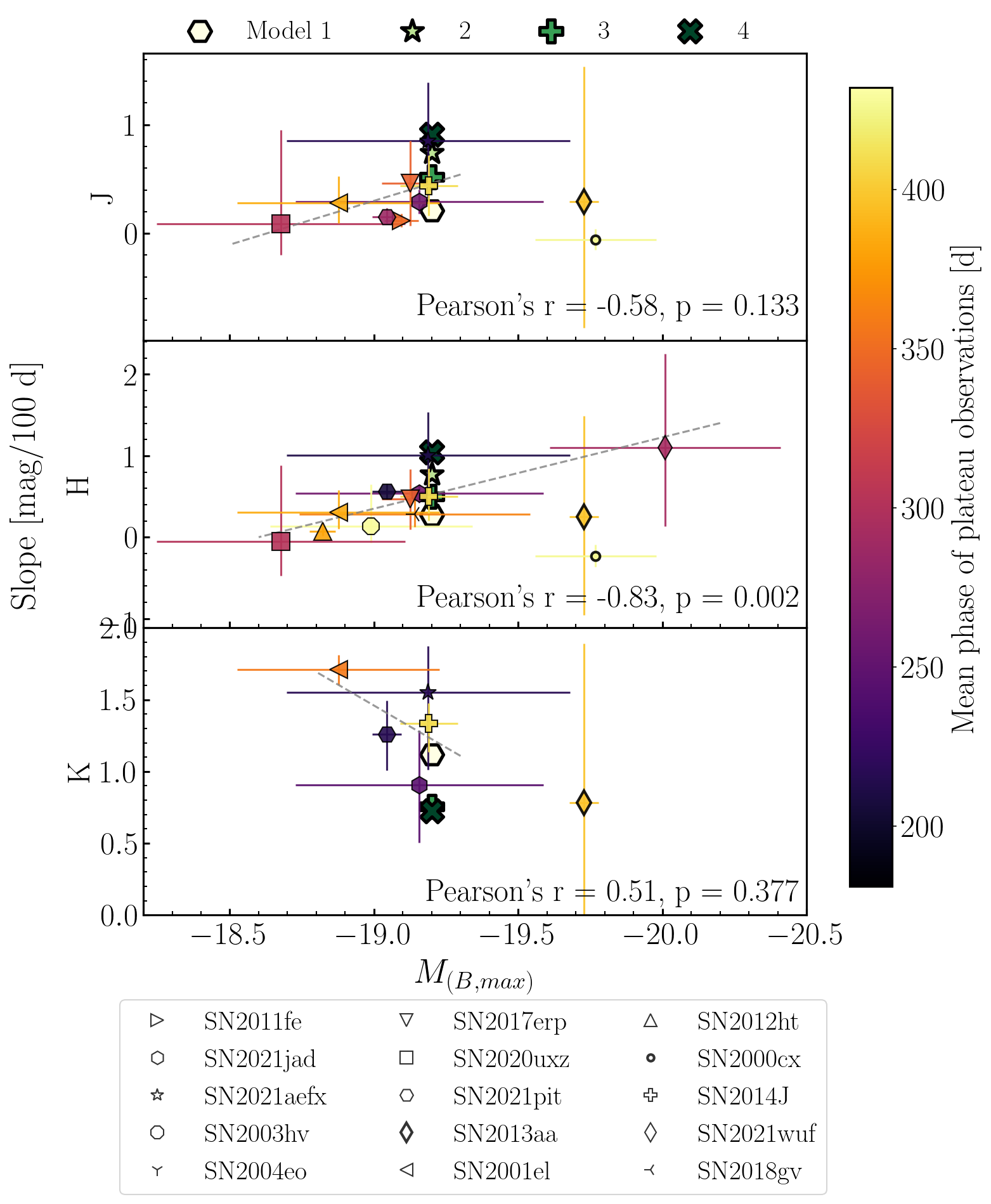}
    \caption{The measured decline rate as a function of \protect$M_{\rm max}^{\rm B}$ for the \textit{J} (top), \textit{H} (middle), and \textit{K$_s$} (bottom) filters. The marker colour represents the mean phase of the observations. We include the predictions from the sub-M$_{\rm ch}$ models presented by \protect\cite{Shingles2022}, using -19.2 as \protect$M_{\rm max}^{\rm B}$ from \protect\cite{Shen2018}. The dashed line shows a linear fit to the sample, excluding SNe 2013aa and 2000cx, which are considered as outliers and are highlighted by a thicker marker edge. We also show the Pearson $r$-coefficient and the corresponding $p$-values. A significant correlation ($p$ < 0.05) is identified in the \textit{H} band.}
    \label{slope_peak_phase}
\end{figure}

In Fig. \ref{slope_peak_phase} we show the decline rate for each individual SN as a function of the absolute \textit{B}-band magnitude at peak ($M_{\rm max}^{\rm B}$) and the mean phase of the observations. In the \textit{J} band, most SNe Ia have a slope consistent with zero. The decline rate averaged across the SNe Ia in the \textit{J} band is 0.4 $\pm$ 0.4 mag~/~100~d, where the uncertainty is the standard deviation weighted by the individual uncertainties. In the \textit{H} band, the average decline rate is 0.5 $\pm$ 0.3 mag~/~100~d. The \textit{K$_s$} band behaves differently from the other two bands, with an average decline rate of 1.3 mag~/~100~d and a weighted standard deviation of 0.3 mag~/~100~d. 

When comparing the decline rates between \textit{J} and \textit{H} for each SN Ia in the sample, they are consistent for six SNe Ia. One SN Ia has a steeper decline in \textit{J}, whilst two have a steeper decline in \textit{H}. Those with a steeper decline in the \textit{H} band have observations limited to <~250 d, whereas those with a consistent decline rate, or shallower decline rate in \textit{H}, have observations taken later than 250~d. This points to an evolution in the decline rate in the \textit{H} band across the plateau, with a steeper intial decline in \textit{H} which levels off with time. This evolution was also apparent in the well-sampled light curves of SN 2017erp and SN 2018gv \citep{Graur2020}, where at the start of the plateau phase the decline rate decreases, but near the end of the plateau phase the decline rate begins to rise again. Unfortunately, no well-sampled \textit{J}-band light curve is available for comparison, but we refer the reader to Appendix \ref{derivatives_analysis} for an analysis of this evolution for model 1.

\subsubsection{Average magnitude during the plateau}\label{ave_mag_text}

\begin{figure}
    \centering
    \includegraphics[width=8.4cm]{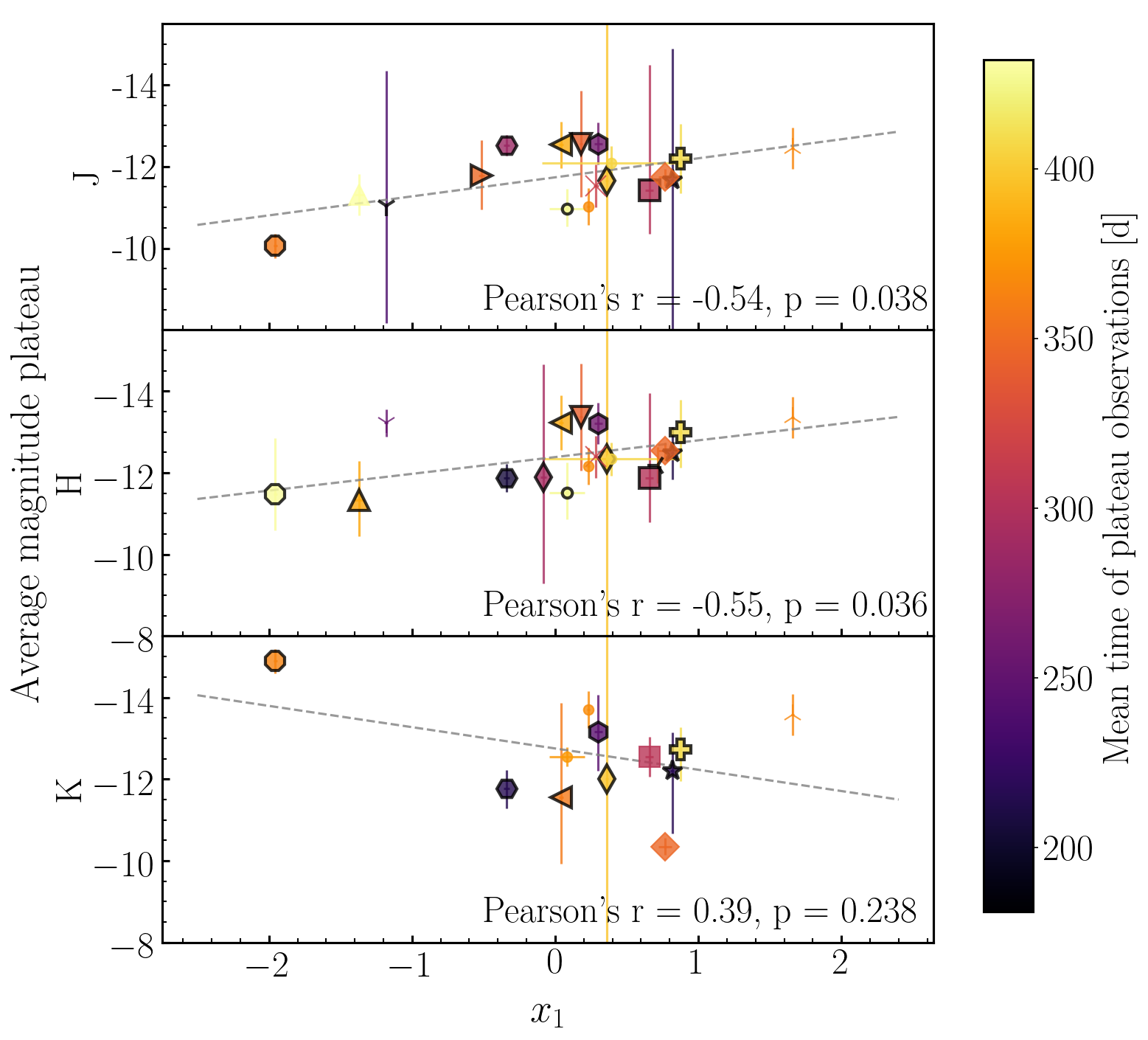}
    \caption{The average magnitude measured for each SN Ia during the plateau as a function of $x_1$, with the colour indicating the mean phase of the observations, for the \textit{J} (top), \textit{H} (middle), and \textit{K$_s$} (bottom) bands. We include measurements taken from a single data point (shown as markers without a black outline), which should give a reasonable estimate of the magnitude in the flatter \textit{J} and \textit{H} bands, but should be interpreted with caution for the \textit{K$_s$} band due to its steeper evolution throughout the plateau. We perform a linear fit (dashed line) in each filter and include the calculated Pearson's $r$-coefficient and the corresponding $p$-value. We find a significant ($p$ < 0.05) trend of the average magnitude during the plateau with $x_1$ in the \textit{J} and \textit{H} bands.}
    \label{ave_mag_x1}
\end{figure}

In Fig. \ref{ave_mag_x1} we show the average magnitude as a function of the light curve stretch ($x_1$), as well as the mean phase during which the data were taken. SNe Ia with only a single data point during the plateau are included but are indicated by markers without a black outline. For the \textit{J} and \textit{H} bands a single data point should give a reliable estimate of the average magnitude during the plateau due to the approximately flat decline rate in these two filters. However, the \textit{K$_s$} band estimates for these SNe Ia are more uncertain due to the steeper decline in this filter.

\subsection{Correlations between plateau properties and SN properties at peak} \label{correlations}

\cite{Graur2020} find that the average magnitude during the plateau in the \textit{H} band scales with $\Delta m_{100}(H)$ (the decrease in magnitude between peak and 100 d after peak in the \textit{H} band) and $\Delta m_{15}(B)$. In the following section we explore the correlations between the plateau properties, $M_{\rm max}^{\rm B}$, $c$, and $x_1$ (available in Table \ref{sample_overview}). 

To measure how strongly two variables are linearly related, we use Pearson's correlation coefficient, $r$. The significance of the correlation is measured by the $p$-value, with $p$<0.05 indicating a statistically significant correlation. We find a significant correlation between $x_1$ and the average \textit{J}- and \textit{H}-band magnitudes during the plateau  ($r$ = $-$0.54, $-$0.55 and $p$ = 0.038 and 0.036, respectively), implying that SNe Ia with broader light curves (larger $x_1$ values) are intrinsically brighter in \textit{J} and \textit{H} during the plateau (Fig. \ref{ave_mag_x1}). This trend agrees with the correlation found by \cite{Graur2020} for the \textit{H} band. We find no statistically significant correlation between the average magnitude during the plateau and $x_1$ in the \textit{K$_s$} band. 

The average magnitude during the plateau is driven predominantly by the slope of the decline prior to the transition onto the plateau, a metric that can be approximated by $\Delta m_{100}(H)$, as shown by \cite{Graur2020}. $\Delta m_{100}(H)$ shows a weak correlation with $\Delta m_{15}(B)$, as shown in fig. 3 of \cite{Graur2020}. Combining these results from \cite{Graur2020} and this work, we suggest that broader SNe Ia (larger $x_1$, smaller $\Delta_{15}(B)$) tend to decline less in the period 100 days after maximum in \textit{H} and therefore have a higher average magnitude during the plateau phase. 

One potential source of bias worth considering is that the likelihood of being able to observe a SN Ia during the plateau is a function of its brightness on the plateau. If a SN Ia has a shallower decline after maximum (smaller $\Delta m_{100}(H)$), it will remain brighter during the plateau. Therefore, it is likely that studies of SNe Ia on the plateau are inherently biased and tend to sample the SNe Ia that are brighter during the plateau and lie at the lower end of the $\Delta m_{100}(H)$ population.

We find no significant correlations between the slope during the plateau and $M_{\rm max}^{\rm B}$, $x_1$, or $c$. However, we note that SNe 2000cx and 2013aa are clear outliers in $M_{\rm max}^{\rm B}$ vs. slope (Fig. \ref{slope_peak_phase}). We check for a linear correlation between the slope and $M_{\rm max}^{\rm B}$ excluding these two SNe Ia. The result from this fit is shown as the dashed line in Fig. \ref{slope_peak_phase}. This correlation is significant in \textit{H} with Pearson's $r$ coefficient = $-$0.83 and $p$-value = 0.002, implying that SNe Ia that are more luminous at peak tend to decline faster during the plateau phase. It is unclear why SNe 2000cx and 2013aa do not follow this trend, but both are very luminous at peak ($M_{\rm max}^{\rm B}$ < -19.5). We discuss these objects in more detail in Section \ref{peculiar_subtypes}. 

The timing of the secondary maximum of SNe Ia shows a strong correlation with the stretch of the light curve \citep{Dhawan2015, Papadogiannakis2019a}, with narrow, fast evolving SNe Ia having an earlier secondary maximum. An increase in the total mass of \nick\ (corresponding to a smaller $\Delta m_{15}(B)$) delays the onset of the secondary maximum due to the higher temperature of the ejecta \citep{Kasen2006}. We suggest that the NIR plateau is caused by a similar mechanism as the secondary maximum, and we expect similar correlations to hold for the NIR plateau. The timing of the onset of the plateau could therefore be expected to correlate with the stretch of the light curve.  We test whether there is any correlation between $x_1$ and the $t_0$ values calculated in Section \ref{onset_fits} and find no statistically significant correlations ($p$-values = 0.8 and 0.1 for the \textit{J} and \textit{H} bands, respectively). However, for most SNe the phase of the onset is very poorly constrained due to poor sampling, and we cannot rule out a possible correlation between these parameters. Future studies of SNe Ia with higher cadence observations (< 20 d) around the transition phase (70 -- 150 d) will help to answer this question.

\section{Discussion} \label{discussion}

In Section \ref{spectral_discussion}, we provide a theoretical discussion of the NIR spectral evolution. We then answer the questions raised by \cite{Graur2020}: ``Is there a plateau in the \textit{K$_s$} band?'' and ``Does the \textit{H}-band plateau consist of two distinct branches?'', by analysing the results presented in Section \ref{results}. We discuss how the models presented in Section \ref{model_description} compare to our observations in Section \ref{compare_models}. In Section \ref{peculiar_subtypes}, we discuss the peculiar SN Ia sub-types present in the sample and compare their behaviour on the plateau to the normal SNe Ia.

\subsection{Relating the photometric evolution to spectral features}\label{spectral_discussion}

\begin{figure*}
    \centering
    \includegraphics[width=17cm]{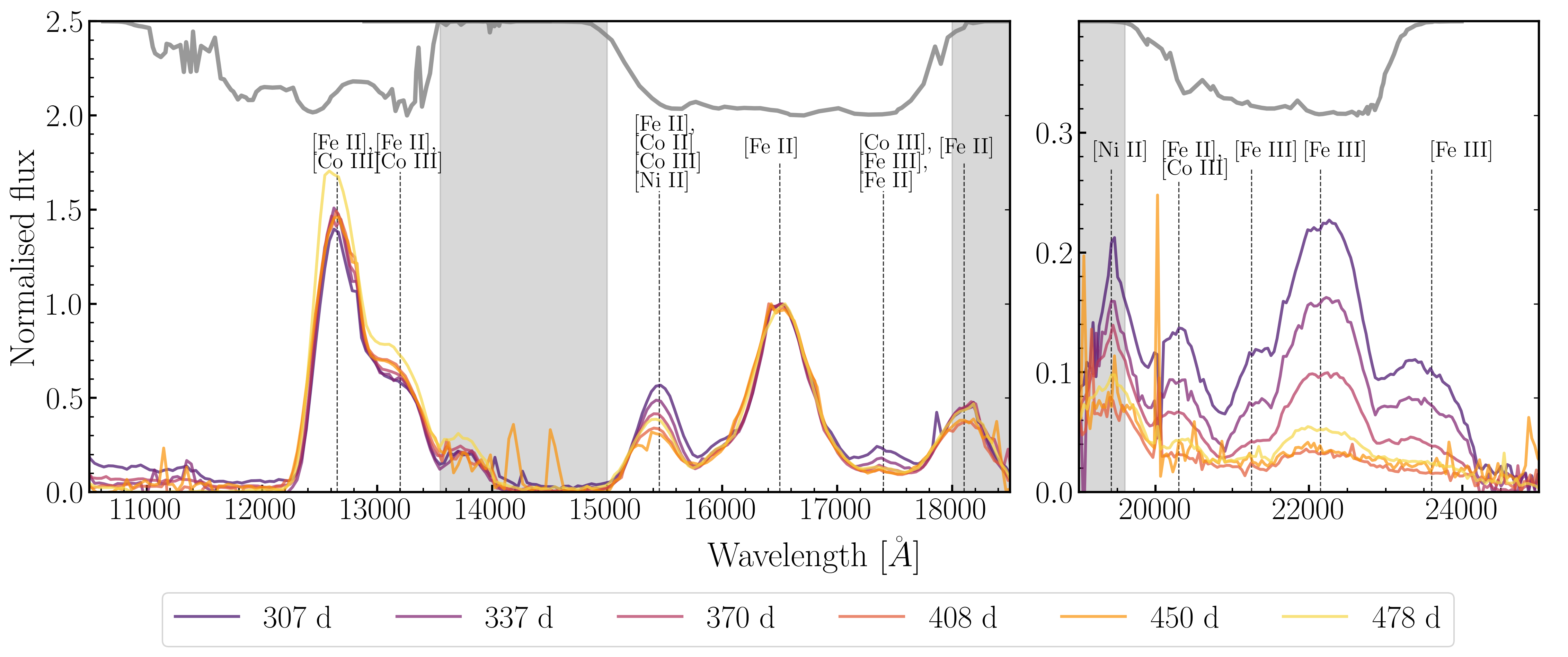}
    \caption{NIR spectral series of SN 2014J, covering 307 -- 478~d. \textit{Left:} The wavelength range covering the \textit{J} and \textit{H} bands. \textit{Right:} The \textit{K$_s$}-band spectrum (note the different range on the y-axis). The key spectral features are marked, and the telluric regions are indicated by grey regions. The flux has been normalised to the \FeIIF\ feature at 1.65 µm. The grey solid curves show the 2MASS \textit{J}, \textit{H}, and \textit{K$_s$} filter response functions.}
    \label{14J_spectra}
\end{figure*}

\begin{figure}
    \centering
    \includegraphics[width=8.4cm]{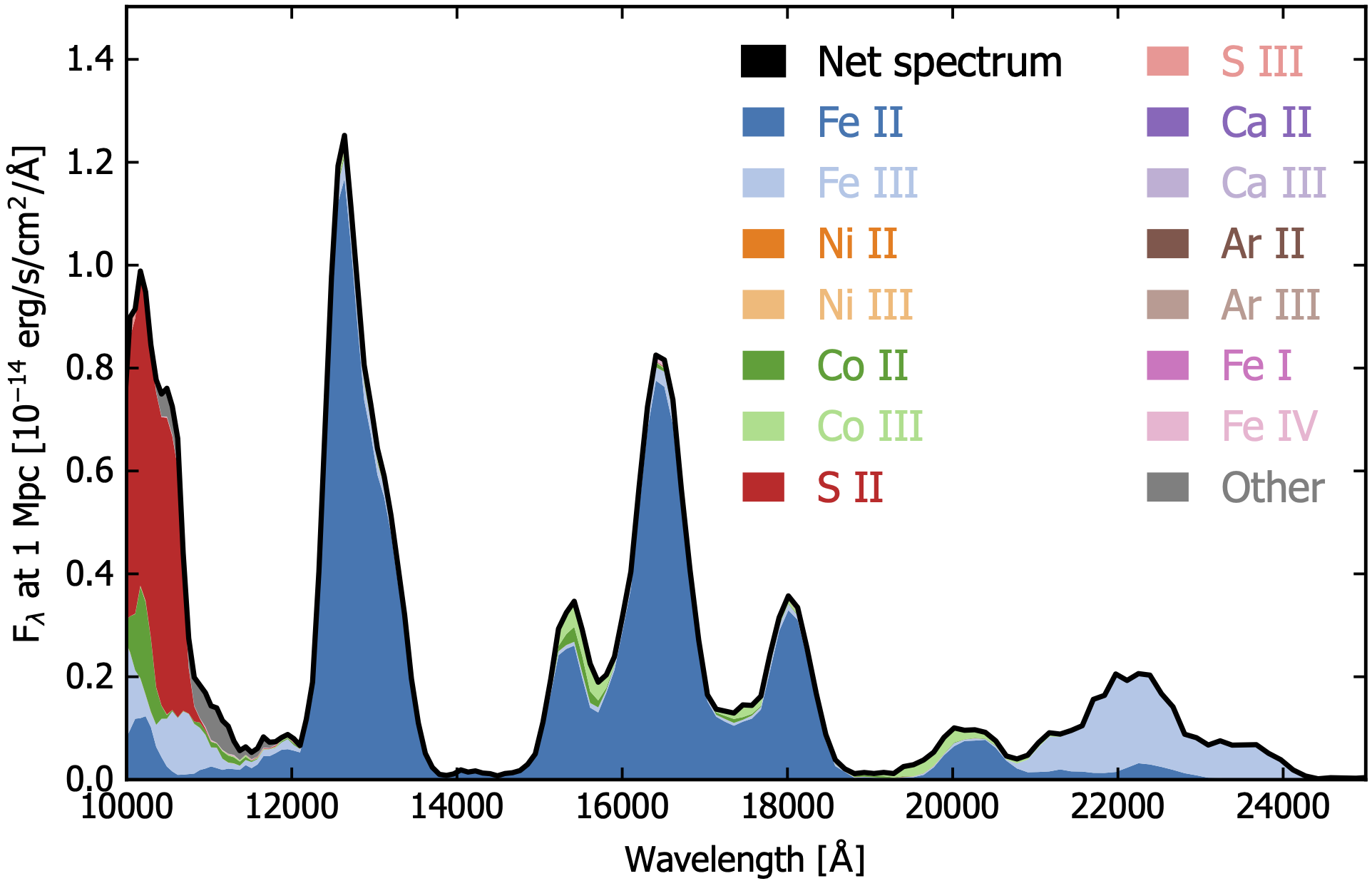}
    \caption{Figure adapted from \protect\cite{Shingles2022} with showing the contribution from each species to the sub-M$_{\rm ch}$ heatboost8 spectrum at 247~d. Most of the \textit{J} and \textit{H} bands are dominated by \FeIIF, with some \CoIIIC\ contributions in the \textit{H} band. The \textit{K$_s$} band is dominated by \FeIIIF\ emission.}
    \label{species_breakdown}
\end{figure}

The NIR spectrum during the plateau contains many forbidden iron group lines. We show the spectral evolution of SN~2014J throughout the plateau in Fig. \ref{14J_spectra}, with the main spectral features indicated (these spectra were previously published by \citealt{Dhawan2018} and \citealt{Diamond2018}). The strength of the lines at 1.54 µm, 1.74 µm (\textit{H} band), 2.02 µm, 2.15 µm, 2.22 µm, and 2.35 µm (\textit{K$_s$} band) decrease with time. These lines contain emission features coming from \CoIIC, \CoIIIC, \FeIIF, and \FeIIIF, although from Fig. \ref{14J_spectra} alone it is not possible to say which emission lines from which elements dominate each feature.

To learn more about the individual contributions to each emission feature, we use information about the transition probability of each line from the atomic data made available by the National Institute of Standards and Technology (NIST).\footnote{https://physics.nist.gov/PhysRefData/ASD} The transitions are optically thin, so their fluxes are proportional to their upper level population times their emission probability. The upper level populations will be similar if their excitation energies are similar, they have similar statistical weights (g=2J+1, where g is the statistical weight and J is the quantum number representing the combined total angular momentum of the electron), and they are both metastable states (only forbidden downward transitions). Therefore, we use the ratios of transition probabilities as a proxy for line strength ratios if the emission lines originate from the same species and have similar excitation energies for the upper level \citep{Jerkstrand2015}. 

The two \CoIIIC\ features in the \textit{J} band (1.27 µm and 1.31 µm) and the two \CoIIIC\ features in the \textit{H} band (1.54 µm and 1.74 µm) all have similar upper energy levels (23 060.95, 23 060.95 , 23 435.93, and 22 721.42 cm$^{-1}$, respectively). The 1.27 µm and 1.31 µm lines originate from the same $a^4P$ multiplet, but come from states with J~=~5/2 and J~=~3/2, respectively, meaning that the 1.27 µm feature is expected to be about (5+1)/(3+1) = 1.5 times stronger. The 1.54 µm and 1.74 µm lines come from the same upper state ($a^{2}G_{9/2}$), so the $A_{\rm ki}$ ratio provides a reliable estimate of the flux ratio of these two lines. The emission line at 1.54 µm has the highest transition probability ($A_{\rm ki}$ = $1.3 \times 10^{-1}$ s$^{-1}$), whereas the lines in the \textit{J} band have transition probabilities of $5.4 \times 10^{-3}$ and $2.5 \times 10^{-3}$ s$^{-1}$, respectively. The line at 1.74 µm has a transition probability of $4.2 \times 10^{-2}$ s$^{-1}$. The dominant \CoIIIC\ features therefore sit in the \textit{H} band, and this band will be most impacted by the decay from \cob\ $\rightarrow$ \fe. 

The \FeIIIF\ features at 2.15, 2.22, and 2.35 µm in the \textit{K$_s$} band have similar upper energy levels (25 142.12, 24 558.44, and 24 558.44  cm$^{-1}$, respectively), all originate from $^3G$, have comparable transition probabilities ($A_{\rm ki}$ = 3.4 $\times 10^{-2}$, 3.20 $\times 10^{-2}$, and 2.25 $\times 10^{-2}$ s$^{-1}$), but come from different states (J~=~4, 6, 5, respectively) meaning that the feature at 2.22 µm is the strongest of the three.

The features in the \textit{J} band show only limited decay with time relative to the \textit{H} band, which aligns with the lower transition probabilities of the \CoIIIC\ features at these wavelengths. This is further supported by Fig. \ref{species_breakdown}, which deconstructs the spectrum of model 1 to show the contributions from different species. This is likely not the only correct model for all SNe Ia, but it demonstrates the commonly identified features whilst also being able to model the plateau behaviour. The model suggests that the \textit{J} band is dominated by \FeIIF, whereas the \textit{H} band has significant contribution from \CoIIIC. The feature at 1.74 µm is composed of three emission lines from \FeIIF, \FeIIIF, and \CoIIIC. The feature at 1.74 µm is dominated by \FeIIF, but the model demonstrates that it also has significant contribution from \CoIIIC, explaining its decay with time. This is in agreement with previously identified features in SNe Ia. The \textit{K$_s$} band is dominated by \FeIIIF\ features (see Fig. \ref{species_breakdown}).

\subsubsection{Is there a plateau in the \textit{K$_s$} band?}

\cite{Graur2020} speculated that based on the synthetic photometry of SN 2014J, the NIR plateau does not extend to the \textit{K$_s$} band. In Section \ref{results}, we presented additional data in the \textit{K$_s$} band supporting this conclusion. Here, we rationalise the lack of a plateau in the \textit{K$_s$} band by referencing the spectroscopic evolution of SN 2014J as a representative of a normal SN Ia (see Fig. \ref{14J_spectra}).

As shown in Figs. \ref{14J_spectra} and \ref{species_breakdown}, the $K$-band is dominated by an \FeIIIF\ complex, whereas the $J$ and $H$ bands are dominated by \FeIIF\ features \citep{Diamond2018, Shingles2022}. Throughout the plateau phase, the strength of \FeIIIF\ features decreases whereas \FeIIF\ features remain constant, suggesting that doubly ionised iron ions are recombining to singly ionised iron. Since there is no contribution of \FeIIF\ in the \textit{K$_s$} band, the flux continues to decline following the recombination rate of \FeIIIF.

\subsubsection{Are there two branches in the \textit{H}-band plateau?}

\cite{Graur2020} find two clusters in the average magnitude of the \textit{H} band, corresponding to a more luminous branch and a faint branch. We note that the photometry in fig. 2 in \cite{Graur2020} is scaled to $M_{\rm max}^H$ whereas the photometry in Fig. \ref{phot_plot} is not scaled because \textit{H}-band data around peak is not available for all the SNe Ia in our sample. A direct comparison between the plots is therefore not possible, but we note that the \textit{H}-band data in this paper is not separated into two different branches. This could mean that the magnitudes of the plateau in the \textit{H} band make up a continuous distribution, but only the extremes of this population were sampled by \cite{Graur2020}. 

\begin{figure}
    \centering
    \includegraphics[width=8.4cm]{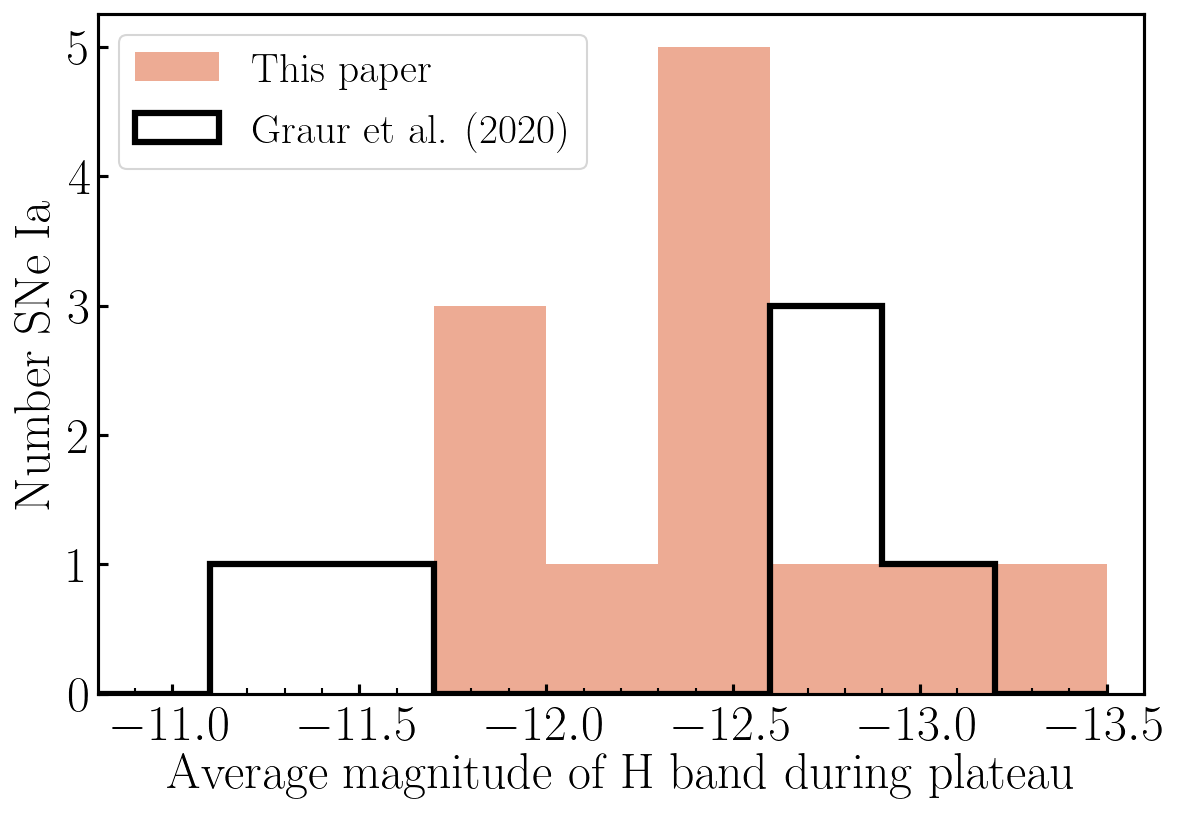}
    \caption{A histogram comparing the average magnitude of the plateau in the \textit{H} band between \protect\cite{Graur2020} and this paper. \protect\cite{Graur2020} noticed bimodal behaviour in the average magnitudes of the \textit{H} band, but the additional data presented in this paper suggests the \textit{H}-band magnitudes represent a continuous distribution.}
    \label{hist_ave_mags}
\end{figure}

We test this first explanation by comparing the average \textit{H}-band magnitudes on the plateau of the SNe Ia presented by \cite{Graur2020} to the additional SNe Ia presented in this paper (Fig. \ref{hist_ave_mags}). The gap between $-$11.5 and $-$12.5 mag found by \cite{Graur2020} is populated by the SNe Ia presented in this paper, suggesting that the \textit{H}-band magnitudes on the plateau represent a continuous distribution. A simple Kolmogorov-Smirnov (KS) test enables us to test whether the two samples are likely sampled from the same distribution. We find a KS-value = 0.4 and $p$-value = 0.5, suggesting that the two samples are most likely drawn from the same population. 

Despite not seeing two separate branches in the \textit{H}-band plateau, the behaviour in the \textit{H} band is less homogeneous than in the \textit{J} band. Fig. \ref{14J_spectra} shows a strong \FeIIF/\CoIIC/\CoIIIC\ complex present in the \textit{H} band, which is dominated by \FeIIF\ and \CoIIIC. This feature decreases in strength throughout the plateau due to the continued decay of $^{56}$Co to $^{56}$Fe \citep{Childress2015, Flors2018}, which would suggest that there should be some decrease in flux in the \textit{H} band during the plateau phase. This aligns well with the model predictions, which suggest that the \textit{J} band decays slower during the plateau phase than the \textit{H} band. We would expect the decline rate in the \textit{H} band to flatten with time, as the relative contribution of the \CoIIIC\ feature decreases and the decay of these features will have a smaller overall impact on the integrated flux across the filter. Generally, the SNe Ia with observations taken at later phases have shallower declines (with the exception of SN 2020uxz), suggesting that the variation seen in the \textit{H} band could be driven by whether the observations are taken during the early or late stages of the plateau.

\subsection{Comparing models to the observations} \label{compare_models}

The light curve of model 1 is shown in Fig. \ref{phot_plot}, where it is scaled to the \textit{J}-band photometry of our sample, the wavelength range best matched by the model presented by \cite{Shingles2022}. In the \textit{H} and \textit{K$_s$} bands, the model under-predicts the magnitude. In the \textit{H} band the discrepancy is greatest near the beginning of the plateau but lessens with time, whereas in the \textit{K$_s$} band the offset remains constant. This mismatch between the relative model flux and observed flux in each band is likely due to specific spectral features not being reproduced as well by the models. Figure \ref{species_breakdown}, which is an extended version of fig. 5 from \cite{Shingles2022}, highlights that the \textit{J} band is dominated by an \FeIIF\ complex spanning 1.22 -- 1.36 µm. The \textit{H} band contains a complex of \FeIIF, \FeIIIF, \CoIIC, and \CoIIIC. Fig. 5 in \cite{Shingles2022}, which compares the model spectrum to the spectrum of SN 2013ct, demonstrates that the model is able to reproduce most spectral features across the \textit{J} and \textit{H} bands. However, the feature at 1.54 µm, which is dominated by \FeIIF\ in Fig. \ref{species_breakdown}, is underestimated. This feature also has contributions from \FeIIIF, \CoIIC, and \CoIIIC, which may be underestimated by model 1.

In Fig. \ref{slope_peak_phase} we show the average decline rate of model 1 on the plateau, between 150--500~d. We find $\sim$0.2, 0.3 and 1.1 mag~/~100~d in the \textit{J}, \textit{H}, and \textit{K$_s$} bands, respectively. This is in agreement with the observational data regarding the presence of the plateau in the \textit{H} and \textit{J} bands, as well as the lack of a plateau in the \textit{K$_s$} band. Moreover, the decline rate predicted by model 1 sits in the parameter space defined by our sample. Fig. \ref{slope_peak_phase} also shows the average decline rates for the other three sub-$M_\mathrm{ch}$ models. All four models fall within the parameter space set by the observed SNe Ia, although in the \textit{J} and \textit{H} bands, models 2, 3, and 4 tend to predict steeper declines than the majority of our sample (with the exception of SN 2021wuf). We provide a more detailed analysis of the magnitude evolution of model 1 in Appendix \ref{derivatives_analysis}, including an analysis of the first and second derivatives, to characterise the evolution of the slope as well as the inflection points.

\subsection{Peculiar SN Ia sub-types on the plateau}\label{peculiar_subtypes}

The majority of the SNe Ia presented in this paper are classified as ``normal'' SNe Ia based on their maximum-light spectra (see Table \ref{sample_overview}), although there are a few exceptions. SN 2021wuf is classified as a 91T-like SN Ia, a subclass that follows the width-luminosity relation \citep{Rust1974, Pskovskii1977, Phillips1993} and is used for cosmology, but with light curves that are generally brighter and slower evolving than normal SNe Ia. They show a preference for exploding in late-type galaxies \citep{Taubenberger2017}. SN 2000cx is a peculiar SN Ia, with properties similar to the 91T-like sub-class but with an asymmetric \textit{B}-band light curve and a peculiar colour evolution \citep{Li2001}. SNe 2004eo and 2012ht are classified as transitional objects between normal and sub-luminous SNe Ia \citep{Yamanaka2014}. Whether these transitional SNe Ia should be used for cosmology is an on-going debate \citep[][Harvey et al. subm.]{Gall2018, Burns2018, Dhawan2022}. 

SNe 2000cx, 2012ht, and 2021wuf all have consistent average magnitudes in \textit{H} during the plateau and follow the trend that narrower SNe Ia tend to be fainter during the plateau. SN 2004eo sits above this trend in \textit{H}, being more luminous than expected for its measured $x_1$. However, in \textit{J} SN 2004eo is consistent with this trend. 

We found a correlation between $M_{\rm max}^{\rm B}$ and the slope, and whilst the low decline rate of SN 2012ht fits into this trend, SN 2000cx is a clear outlier (Fig. \ref{slope_peak_phase}). SN 2013aa, whilst being classified as a normal SN Ia, is also exceptionally luminous and similarly falls outside this correlation. On the other hand, SN 2021wuf shows a steep decline during the plateau, as expected from the correlation, although we note that this measurement is based on only two data points separated by 26 d (a minimum of 25 d is required to calculate a reliable slope). SN 2004eo only has two data points in \textit{J}, and these are not sufficiently spaced to calculate a decline rate. 

A larger sample is required to investigate these trends, but if over-luminous SNe Ia tend to have a flatter plateau in \textit{J} and \textit{H}, this may imply that there is an additional spectral contribution at these wavelengths supporting their luminosity for a longer period. 

It has been suggested that the single-degenerate scenario with a near-M$_{\rm ch}$ WD could be solely responsible for over-luminous 91T-like SNe Ia rather than the normal SN Ia population \citep{Fisher2015, Byrohl2019, Childress2015}. Previous studies have also found that over-luminous 91T-like SNe Ia show flux excesses at a higher rate than normal SNe Ia, which could point towards interaction with a non-degenerate companion in the single-degenerate scenario (\citealt{Jiang2018, Deckers2022} but see \citealt{Burke2022b} for an alternative view).

\section{Conclusions}\label{conclusion}

We present NIR photometry of 24 SNe Ia during the plateau phase. From this extensive data set we are able to measure the average magnitude and slope of the plateau in \textit{J}, \textit{H}, and \textit{K$_s$}. We compare these plateau properties to the properties at maximum light and find a significant correlation between $x_1$ and the magnitude of the plateau in \textit{J} and \textit{H}, as well as between $M_{\rm max}^{\rm B}$ and the slope in \textit{H}. From these correlations we conclude that the main driving factor for the magnitude of the plateau is the luminosity at maximum light, which in turn correlates with the decline in magnitude in \textit{H} 100 d after maximum, ($\Delta m_{100}(H)$). SNe Ia which are more luminous at peak appear to decline faster during the plateau, although there are clear outliers to this trend. Specifically, the over-luminous SNe in our sample behave differently from the normal SNe Ia during the plateau. Over-luminous SNe Ia appear to decline slower than predicted by the trend found between $M_{\rm max}^{\rm B}$ and the slope, which could imply that there is an additional spectral contribution during the plateau.

We constrain the onset of the plateau to 70 -- 150~d. The secondary maximum occurs in \textit{H} before it occurs in \textit{J} \citep{Kasen2006, Dhawan2015}, but due to the large uncertainties in our estimates of the transition phase we are unable to determine if this is the case for the plateau. We expect a correlation to exist between the time of the onset of the plateau and the peak luminosity of a SN Ia, akin to the correlation found for the secondary maximum, but this could not be confirmed for our sample. 

We compare our photometry to models produced by \cite{Shingles2022} and find good agreement regarding the evolution during the plateau, albeit the models under-predict the luminosity in \textit{H} and \textit{K$_s$}. However, the best-matching model has reduced non-thermal ionisation rates which leads to lower ionisation states, but no physical justification for reducing these rates has yet been proposed.

An analysis of six spectra of SN 2014J taken throughout the plateau enables us to explain the presence of the plateau in \textit{J} and \textit{H}, as well as the absence of the plateau in \textit{K$_s$}. The dominant \FeIIF\ features which remain constant throughout the plateau sit in the \textit{J} and \textit{H} bands, whilst the \textit{K$_s$} band hosts mainly \FeIIIF\ features, which recombine to \FeIIF\ during the plateau phase.

A very limited number of SNe Ia have NIR coverage during the onset of the plateau. Extending this parameter space by obtaining higher cadence observations (< 20 d) around the transition phase (70 -- 150 d) will enable us to test whether the timing of the plateau correlates with the magnitude at peak, as is the case for the secondary maximum, although we note that this is often difficult due to visibility constraints from the ground. We strongly encourage follow up of over-luminous SNe Ia to test whether they all decline faster during the plateau than expected, since this might imply these events have a different origin. Finally, obtaining more UV photometry coeval with NIR photometry would enable us to determine if flux truly is being redistributed from the UV to the NIR.

\section*{Acknowledgements}
The authors would like to thank the anonymous referee for helpful comments that have improved this paper. MD extends her gratitude to the Institute of Cosmology and Gravitation at the University of Portsmouth for hosting them throughout this work. MD and KM are funded by the EU H2020 ERC grant no.758638. LJS acknowledges support by the European Research Council (ERC) under the European Union’s Horizon 2020 research and innovation program (ERC Advanced Grant KILONOVA No. 885281). SJB would like to thank their support from Science Foundation Ireland and the Royal Society (RS-EA/3471). L.G. acknowledges financial support from the Spanish Ministerio de Ciencia e Innovación (MCIN), the Agencia Estatal de Investigación (AEI) 10.13039/501100011033, and the European Social Fund (ESF) "Investing in your future" under the 2019 Ramón y Cajal program RYC2019-027683-I and the PID2020-115253GA-I00 HOSTFLOWS project, from Centro Superior de Investigaciones Científicas (CSIC) under the PIE 20215AT016 and LINKA20409 projects, and the program Unidad de Excelencia María de Maeztu CEX2020-001058-M. TEMB acknowledges financial support from the Spanish Ministerio de Ciencia e Innovación (MCIN), the Agencia Estatal de Investigación (AEI) 10.13039/501100011033 under the PID2020-115253GA-I00 HOSTFLOWS project, from Centro Superior de Investigaciones Científicas (CSIC) under the PIE project 20215AT016 and the I-LINK 2021 LINKA20409, and the program Unidad de Excelencia María de Maeztu CEX2020-001058-M. MN is supported by the European Research Council (ERC) under the European Union’s Horizon 2020 research and innovation programme (grant agreement No.~948381) and by a Fellowship from the Alan Turing Institute. This work makes use of data from Las Cumbres Observatory. The LCO group is supported by NSF grants AST-1911225 and AST-1911151, HST-GO-16497 and HST-GO-16884. This work is based in part on data obtained with the NASA/ESA \textit{Hubble Space Telescope} through programs GO-16497 and 16885. This work is also based on observations collected at the European Organisation for Astronomical Research in the Southern Hemisphere, Chile, as part of ePESSTO+ (the advanced Public ESO Spectroscopic Survey for Transient Objects Survey). ePESSTO+ observations were obtained under ESO program IDs 1103.D-0328, 106.216C, 108.220C (PI: Inserra). This work is also based on observations collected at the European Southern Observatory under ESO programmes 091.D-0764(A), 092.D-0632(A), 099.D-0683(A), 0100.D-0242(A). This research has also made use of NASA's Astrophysics Data System and the NASA/IPAC Extragalactic Database (NED), which is operated by the Jet Propulsion Laboratory, California Institute of Technology, under contract with NASA. This work was funded by ANID, Millennium Science Initiative, ICN12\_009. 
\section*{Data Availability}

All the photometry presented in this paper is made available in machine-readable format in the supplementary material.



\bibliographystyle{mnras}
\bibliography{astro} 




\appendix
\section{Sample overview}

\begin{landscape}
\begin{table}
\centering
\begin{threeparttable}
\caption{A table summarising the SNe Ia present in our sample, showing the classification of the SN, redshift, distance modulus, and time of maximum light. We also include optical and cosmological parameters ($M_{\rm max}^{\rm B}$, $x_1$, and $c$) and the source of the NIR photometry/spectroscopy. }\label{sample_overview}
\begin{tabular}{|l|l|c|c|c|c|c|c|l|l|}
\hline
     \textbf{SN} & \textbf{$z$} & $\mu$ &$t_{\rm max}^{\rm B}$ & $E(B-V)_{\rm gal}$ & $M_{\rm max}^{\rm B}$  & $x_1$ &  $c$ & \textbf{Data Ref.$^{\rm a}$} & \textbf{Dist. ref.$^{\rm b}$} \\
       &  & [mag] & [d]  & [mag] & [mag] &  &   &  & \\
     \hline
SN2000cx$^{\rm c}$   & 0.00818  & 32.87 $\pm$ 0.21  & 51752.1 & 0.082 & $-$19.77 $\pm$ 0.21 & 0.08 $\pm$ 0.12 & 0.043  $\pm$ 0.003 & {[}1{]}, [2]  &    [24]    \\
SN2001el   & 0.003896 & 31.65 $\pm$ 0.35  & 52181.9 & 0.014 & $-$18.88 $\pm$ 0.35 & 0.04  $\pm$ 0.02 & 0.191  $\pm$ 0.002 & {[}3{]}, [4]  &    [25]    \\
SN2003hv   & 0.005624 & 31.37 $\pm$ 0.30   & 52891.7  & 0.016 & $-$18.99 $\pm$ 0.35 & $-$1.95 $\pm$ 0.01 & $-$0.110 $\pm$ 0.002 &  [5] &    [26], [27]     \\
SN2004eo$^{\rm d}$    & 0.016    & 34.12 $\pm$ 0.10   & 53278.1  & 0.108 & $-$19.06 $\pm$ 0.10 & $-$1.18 $\pm$ 0.01 & 0.169  $\pm$ 0.001 & {[}6{]} &   [28]   \\
SN2011fe   & 0.0006  & 29.04 $\pm$ 0.05   & 55814.6  &  0.009 &  $-$19.09 $\pm$ 0.06 & $-$0.51 $\pm$ 0.02 & $-$0.022 $\pm$ 0.003  &  [7], [8] &  [29]  \\
SN2012cg$^{*}$    & 0.0015   & 30.83 $\pm$ 0.05  & 56082.2  & 0.200 & $-$19.55 $\pm$ 0.05 & 0.77  $\pm$ 0.05 & 0.071  $\pm$ 0.005 & {[}9{]}, [10] &     [30]     \\
SN2012fr   & 0.005    & 31.38 $\pm$ 0.06  & 56241.9 & 0.020 & $-$19.45 $\pm$ 0.07 & 1.66  $\pm$ 0.01 & $-$0.084 $\pm$ 0.001 & {[}11{]}, [13]&  [31]\\
SN2012ht$^{\rm e}$   & 0.003    & 31.91 $\pm$ 0.04  & 56295.1  & 0.029 & $-$18.30 $\pm$ 0.40 & $-$1.37 $\pm$ 0.03 & 0.120 $\pm$ 0.005 & {[}11{]}, [12], [14]  &     [32]    \\
SN2013aa   & 0.003    & 30.72 $\pm$ 0.05  & 56342.5   &0.170 & $-$19.73 $\pm$ 0.05 & 0.36  $\pm$ 0.01 & 0.026 $\pm$ 0.001 & {[}9{]}, [15]      &   [33] \\
SN2013cs   & 0.009    & 32.94 $\pm$ 0.14  & 56437.2   & 0.093& $-$19.33 $\pm$ 0.14 & 0.29  $\pm$ 0.02 & 0.116  $\pm$ 0.002 & {[}9{]} &  [34]\\
SN2013ct   & 0.003    & 30.27 $\pm$ 0.20   & 56416.1 & 0.028 &        $-$ $\pm$ $-$    &       $-$ $\pm$ $-$     &        $-$ $\pm$ $-$      & {[}9{]}   &     [34]   \\
SN2013dy   & 0.003889 & 31.54 $\pm$ 0.08  & 56501.7  &0.350 & $-$19.70 $\pm$ 0.08 & 0.92  $\pm$ 0.02 & 0.283  $\pm$ 0.002 & {[}9{]}, {[}12{]}, [16] &  [32]\\
SN2014J    & 0.000677 & 27.74 $\pm$ 0.08  & 56689.0   &0.050 & $-$19.19 $\pm$ 0.10 & 0.87  $\pm$ 0.01 & 1.198  $\pm$ 0.001 & {[}12{]}, [17]  &     [35]   \\
SN2016hvl  & 0.0131   & 33.76 $\pm$ 0.40  & 57710.9  & 0.438 &         $-$18.40 $\pm$ 0.40    &      0.23 $\pm$ 0.01      &        0.458 $\pm$ 0.001      & {[}18{]} &    [36]      \\
SN2017cbv$^{*}$  & 0.003    & 31.14 $\pm$ 0.40   & 57841.3  & 0.169 & $-$20.13 $\pm$ 0.40 & 0.62  $\pm$ 0.01 & 0.082  $\pm$ 0.001 & {[}18{]}, [19] &     [37]     \\
SN2017erp  & 0.006174 & 32.34 $\pm$ 0.10 & 57934.6  & 0.100 & $-$19.12 $\pm$ 0.10 & 0.18  $\pm$ 0.01 & 0.175  $\pm$ 0.001 & {[}9{]}, [12], {[}20{]}, [21] & [38] \\
SN2018gv   & 0.005274 & 31.71 $\pm$ 0.40   & 58149.6   & 0.058 & $-$19.14 $\pm$ 0.40 & 0.717  $\pm$ 0.002 & $-$0.0059 $\pm$ 0.0002 & {[}12{]}, [22]  &      [36]  \\
SN2019np$^{*}$   & 0.00452  & 32.87 $\pm$ 0.43  & 58507.6  & 0.020& $-$19.57 $\pm$ 0.43 & $-$0.86 $\pm$ 0.03 & $-$0.013 $\pm$ 0.002 & {[}12{]}, [23]   &   [39]     \\
SN2020ees  & 0.024424 & 35.16 $\pm$ 0.40   & 58926.1  & 0.014 &       $-$19.10 $\pm$ 0.45     &      0.39 $\pm$  0.48     &         0.024 $\pm$ 0.060      & {[}18{]}  &    [36]     \\
SN2020uxz$^{**}$  & 0.00825  & 32.10  $\pm$ 0.43  & 59143.0  & 0.038 & $-$18.68 $\pm$ 0.43 & 0.66  $\pm$ 0.03 & 0.005 $\pm$  0.004 & {[}18{]}  &     [39]    \\
SN2021jad  & 0.005534 & 31.87 $\pm$ 0.43  & 59328.7   & 0.035& $-$19.16 $\pm$ 0.43 & 0.30  $\pm$ 0.03 & 0.028  $\pm$ 0.003 & {[}18{]} &      [39]    \\
SN2021pit  & 0.004    & 31.31 $\pm$ 0.05  & 59384.5   & 0.014 & $-$19.04 $\pm$ 0.05 & $-$0.34 $\pm$ 0.02 & 0.027  $\pm$ 0.002 & {[}18{]}  &      [32]   \\
SN2021wuf  & 0.01     & 33.61 $\pm$ 0.40   & 59461.1   & 0.090 & $-$20.01 $\pm$ 0.40 & $-$0.08 $\pm$ 0.02 & $-$0.136 $\pm$ 0.002 & {[}18{]}   &    [36]    \\
SN2021aefx$^{*}$ & 0.005017 & 31.27 $\pm$ 0.49  & 59545.9  & 0.009 & $-$19.19 $\pm$ 0.49 & 0.82  $\pm$ 0.01 & 0.002 $\pm$ 0.001 & {[}18{]}&  [40]\\
    \hline
\end{tabular}
  \begin{tablenotes}
    \footnotesize 
        \item[\rm a] Sources of photometry and spectroscopy: (1) \cite{Sollerman2004}, (2) \cite{Li2001}, (3) \cite{Stritzinger2007}, (4) \cite{Krisciunas2003}, (5) \cite{Leloudas2009}, (6) \cite{Pastorello2007}, (7) \cite{Shappee2017}, (8) \cite{Vinko2012}, (9) \cite{Maguire2016}, (10) \cite{Marion2016}, (11) \cite{Maguire2013}, (12) \cite{Graur2020}, (13) \cite{Zhang2014a}, (14) \cite{Vinko2018}, (15) \cite{Burns2020}, (16) \cite{Zhai2016}, (17) \cite{Sand2016}, (18) this paper, (19) \cite{Hosseinzadeh2017}, (20) \cite{Clark2021}, (21) \cite{Brown2019}, (22) \cite{Yang2020a}, (23) \cite{Sai2022}.
        \item[\rm b] Sources for redshift-indepedent distance moduli: (24) \cite{Li2001}, (25) \cite{Krisciunas2003}, (26) \cite{Tonry2001}, (27) \cite{Jensen2003}, (28) \cite{Shappee2011}, (29) \cite{Pastorello2007}, (30) \cite{Vinko2018}, (31) \cite{Tully2009}, (32) \cite{Riess2016}, (33) \cite{Jacobson-Galan2018}, (34) \cite{Walker2015}, (35) \cite{Tully2013}, (36) \cite{Theureau2007}, (37) \cite{Konyves-Toth2020}, (38) \cite{Tully1988}, (39) \cite{Sorce2014}, (40) \cite{Sabbi2018}.
        \item[\rm c] SN 2000cx is spectroscopically classified as a 91T-like SN Ia, but it shows unusual asymmetry in its \textit{B}-band light curve and its spectral evolution is unlike that of the rest of the subclass \citep{Li2001}. These peculiarities have been attributed to a larger \nick\ mass and higher kinetic energy in the explosion \citep{Li2001}.
        \item[\rm d] SN 2004eo is likely a transitional objects sharing both properties with normal and sub-luminous SNe Ia \citep{Pastorello2007}.
        \item[\rm e] SN 20012ht is also a transitional SN Ia, with properties similar to SN 2004eo \citep{Yamanaka2014}.
        \item[*] A flux excess has been detected in the early light curve.
        \item[**] We tentatively find a flux excess in the SWIFT data for this light curve, see discussion in Section \ref{peculiar_subtypes}.

    \end{tablenotes}
\end{threeparttable}
\end{table}
\end{landscape}

\section{Use of derivatives to analyse the evolution during the plateau}\label{derivatives_analysis}

\begin{figure*}
    \centering
    \includegraphics[width=17cm]{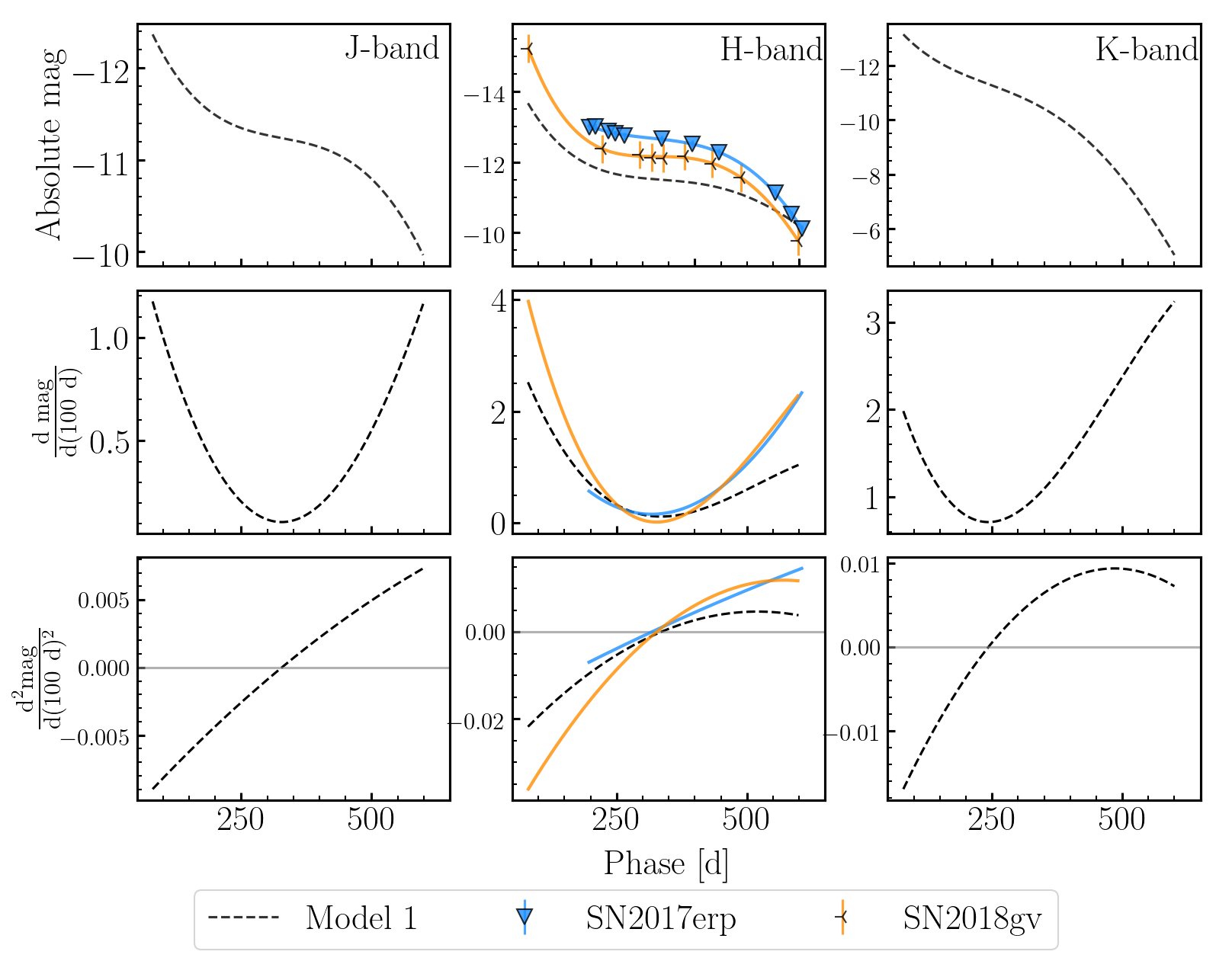}
    \caption{\textit{Top panels:} The light curves in \textit{J}, \textit{H}, and \textit{K$_s$} bands of model 1, as well as SNe 2017erp and 2018gv. All light curves are fit using a univariate spline, and the resulting fits are plotted as a dashed line for the model and solid lines for the SNe Ia. \textit{Middle panels:} The first derivative of the apparent magnitude with respect to time. \textit{Bottom panels:} The second derivative of the apparent magnitude with respect to time. We denote where the second derivative equals zero with a grey line, since where this line meets the second derivative indicates the inflection point in the light curve. Model 1 and SN 2017erp both reach an inflection point at 317 d, whilst SN 2018gv reaches an inflection point at 327~d.}
    \label{model_slope_analysis}
\end{figure*}

It is clear from Fig. \ref{slope_two_component} that there is an evolution in the decline rate for model 1, and finding the average decline rate across the plateau ignores such evolution. To analyse this evolution, we fit the models with a univariate spline and calculate the first and second order derivatives with respect to the phase, presented in Fig. \ref{model_slope_analysis}. Although the \textit{J} and \textit{H} bands both approach a slope of zero around 300 d, the slope in \textit{H} is steeper prior to this and there is a larger change in the first derivative between 150 -- 300 d than in the \textit{J} band. After the minimum is reached, the shape of the first derivative is similar between the two bands. 

The only SNe Ia with sufficient data to capture the full evolution across the plateau are SNe 2017erp and 2018gv. SN 2014J has data spanning 350 -- 500 d, but we exclude it here because the photometry is synthetic and shows unusual evolution, although it is consistent with the evolution of SNe 2017erp and 2018gv within its uncertainties. We include the light curves of SNe 2017erp and 2018gv, the spline fits, and derivatives in Fig. \ref{model_slope_analysis}. The first derivative of the light curve of SN 2017erp matches exceptionally well with the model, whilst SN 2018gv reaches its minimum in the first derivative slightly later. Similarly, the inflection points in the light curves, located where the second derivative equals zero, occur at the same time for the model and SN 2017erp, whereas SN 2018gv reaches the inflection point 10 d later. We note that when measuring the decline rate for the remainder of the sample (see Fig. \ref{slope_peak_phase}), those with data at a later stage (200+ d) have a shallower decline as discussed in Section \ref{decline_results}.


\bsp	
\label{lastpage}
\end{document}